\documentclass[10pt]{article}

\usepackage{fullpage}
\usepackage{setspace}
\usepackage{parskip}
\usepackage{titlesec}
\usepackage[section]{placeins}
\usepackage{xcolor}
\usepackage{breakcites}
\usepackage{lineno}
\usepackage{hyphenat}
\usepackage{setspace}
\usepackage{moresize}
\usepackage{graphicx}
\usepackage{caption}
\usepackage{float}
\usepackage{arydshln}
\usepackage{lineno}

\PassOptionsToPackage{hyphens}{url}
\usepackage[colorlinks = true,
            linkcolor = blue,
            urlcolor  = blue,
            citecolor = blue,
            anchorcolor = blue]{hyperref}
\usepackage{etoolbox}


\usepackage{natbib}
\title{Bibliography management: \texttt{natbib} package}

\titlespacing{\section}{0pt}{*3}{*1}
\titlespacing{\subsection}{0pt}{*2}{*0.5}
\titlespacing{\subsubsection}{0pt}{*1.5}{0pt}

\usepackage{authblk}
\usepackage{graphicx}
\usepackage[space]{grffile}
\usepackage{latexsym}
\usepackage{textcomp}
\usepackage{longtable}
\usepackage{tabulary}
\usepackage{booktabs,array,multirow}
\usepackage{amsfonts,amsmath,amssymb}
\providecommand\citet{\cite}
\providecommand\citep{\cite}

\newif\iflatexml\latexmlfalse

\AtBeginDocument{\DeclareGraphicsExtensions{.pdf,.PDF,.eps,.EPS,.png,.PNG,.tif,.TIF,.jpg,.JPG,.jpeg,.JPEG}}

\usepackage[utf8]{inputenc}
\usepackage[ngerman,greek,english]{babel}

\usepackage{float}

\SetSinglespace{2}
\begin{document}

\title{\textbf{Benefits of Deterministic and Stochastic Tendency Adjustments in a
Climate Model}}

\author[1]{\textbf{William Eric Chapman}}%
\author[1]{\textbf{Judith Berner}}%
\affil[1]{National Center for Atmospheric Research, Boulder,
Colorado,USA}%

\normalsize

\vspace{-1em}

\date{\today}

\begingroup
\let\center\flushleft
\let\endcenter\endflushleft
\maketitle
\endgroup

\sloppy

\noindent\rule{\textwidth}{1pt}

\emph{Corresponding author} : William E. Chapman, wchapman@ucar.edu

\noindent\rule{\textwidth}{1pt}

\Large
\textbf{ABSTRACT}

\normalsize
We develop and compare model-error representation schemes derived from data assimilation increments and nudging tendencies in multi-decadal simulations of the community atmosphere model, version 6. Each scheme applies a bias correction during simulation run-time to the zonal and meridional winds. We quantify to which extent such online adjustment schemes improve the model climatology and variability on daily to seasonal timescales. Generally, we observe a ca. 30\% improvement to annual upper-level zonal winds, with largest improvements in boreal spring (ca. 35\%) and winter (ca. 47\%). Despite only adjusting the wind fields, we additionally observe a ca. 20\% improvement to annual precipitation over land, with the largest improvements in boreal fall (ca. 36\%) and winter (ca. 25\%), and a ca. 50\% improvement to annual sea level pressure, globally. With mean state adjustments alone, the dominant pattern of boreal low-frequency variability over the Atlantic (the North Atlantic Oscillation) is significantly improved. Additional stochasticity further increases the modal explained variances, which brings it closer to the observed value. A streamfunction tendency decomposition reveals that the improvement is due to an adjustment to the high- and low-frequency eddy-eddy interaction terms. In the Pacific, the mean state adjustment alone led to an erroneous deepening of the Aleutian low, but this was remedied with the addition of stochastically selected tendencies. Finally, from a practical standpoint, we discuss the performance of using data assimilation increments versus nudging tendencies for an online model-error representation.

\section{Introduction}
\begin{spacing}{1.1}
In recent decades, significant strides have been made within the scientific community to enhance the numerical modeling of Earth's climate using General Circulation Models (GCMs). These models can emulate atmospheric and oceanic dynamics on various temporal and spatial scales and have broadened our knowledge of the past, present, and future climate system. This progress has led to more precise simulations of weather, extreme events, and climate processes. However, accurately representing all the dynamics that affect the Earth's climate remains a challenge despite these advancements. Many atmospheric processes occur at scales too small for the current resolution of GCMs, and it is unlikely that we will achieve the computational power to represent them soon. Hence, the parameterization of these phenomena continues to be a persistent necessity.

Approximations of the equations of fluid motion through physical parameterization schemes and truncation errors in the numerical discretization give rise to inherent biases within GCMs \citep[e.g.,][]{Berner2012,flato2013}. Consequently, GCM data may not accurately reflect real-world observations and may misrepresent climate and weather risks. A portion of this bias is the result of so-called ``fast-physics'' errors, which manifest in the first few days of a model run as a result of parameterization error \citep[][]{Hurrell2009, Ma2014, Palmer2011, Williamson2007}. These errors may initially be local but can influence model bias at climate timescales by cascading upscale \citep[][]{Jung2008,Klinker1992,Rodwell2007}. Addressing model biases through model physics/parameterization improvement can be challenging, and compensation errors due to model tuning are likely to arise. Thus, some studies have proposed empirically determined corrections to the model's tendency equations. One of these approaches entails utilizing a Data Assimilation (DA) system to learn systematic model bias.

The goal of the current study is to compare two data assimilation (DA) based analysis adjustment techniques for assessing and addressing model error via examining and then re-inserting adjustment increments during model runtime: 1) Relaxation (or nudging) towards observed data and 2) an Ensemble Adjustment Kalman Filter \citep[EAKF,][]{Anderson2001}.

\subsection{Nudging}

Nudging techniques were developed in the late 1980s to study potential remote origins of forecast error by gradually relaxing the model towards analysis (or reanalysis) data during the integration process \citep[e.g.,][]{Ferranti1990,Haseler1982,Klinker1990}. Due to the availability of ever-improving reanalysis data, these techniques have been used to force numerical simulations toward specific observed trajectories.

Researchers have employed nudging to gain insight into the sources of forecast errors and to improve the accuracy of climate models \citep[][]{Ferranti1990,Jung2008,Jung2010} \citep[][]{Ferranti1990,batte2016_new,Jung2008,Jung2011}. Nudging is easy to implement and adds computationally negligible expense to the model simulations. However, these advantages come with some significant disadvantages: 1) the choice of nudging parameters is generally not dynamically driven and is largely subjective; 2) relaxation terms can provide a significant contribution to the tendencies and might have adverse effects on model dynamics; and 3) model dynamics are excluded during the relaxation implementation.

\subsection{Data Assimilation}

An alternative approach is to estimate model bias utilizing a DA system, where the large-scale circulation is initially constrained to match observational data. This step helps eliminate errors that may result from slow atmospheric adjustments to nonlocal processes \citep[e.g.,][]{Karspeck2018,Raeder2012,Raeder2021}. By analyzing the adjustments made to the constrained tendencies (the difference between the model prior and posterior state, referred to henceforth as model increments), we can effectively identify systematic fast-physics errors. Further, these systematic increments can form the basis function of an empirically driven online tendency adjustment which acts to constrain the fast-physics error, and improve mean/variability climate biases \citep[e.g.,][]{Palmer2011}. DA increments have long been leveraged to reveal physical processes in a model which are behaving non-physically \citep[e.g.,][]{Klinker1992,Mapes2012,Rodwell2007,Simpson2018}. The increments, when sufficiently averaged, reveal systematic fast-physics tendencies which indicate erroneous model behavior. \citet{Jung2011} argued, through comparing analysis increments and nudging tendencies in a weather forecasting application, that some of the deficiencies of the nudging technique are ameliorated via the sophistication of a state-of-the-art DA system as there are fewer subjective choices of parameters to be made and spurious model tendencies are reduced. However, the algorithmic sophistication comes with some computational cost, and thus the ability to nimbly rerun, adjust, and evaluate parameter choices in the DA system is rendered difficult.

\subsection{Online Bias Correction}

Whether based on nudging or DA, inserting tendency correction during runtime has an advantage over post-processing methods, which correct archived simulations based on reforecast data \citep[e.g.,][]{Chapman2022,glahn1972use}. Because they correct the atmospheric state only a posteriori, they cannot trigger atmospheric behavior that would have occurred given a correct base state. For example, if sea surface temperatures (SST) are systematically too low near the maritime continent, they cannot give rise to local convection leading to downstream teleconnections which then influence weather in a given hemisphere. Offline bias correction is incapable of remedying this problem, whereas online bias correction can adjust the actual model attractor and allow the model to access observed modes of variability it would have otherwise bypassed. Averaging and re-inserting the increments from nudging or DA during run-time have been previously implemented and can dramatically improve the background model state for weather and climate applications \citep[e.g.,][]{batte2016_new,Chang2019,Crawford2020,Lu2020}. These studies have shown that online corrections result in significant improvements in the skill of weather forecasts which out-compete results obtained by posteriori corrections.

A separate question is if tendency corrections should be inserted deterministically or if random model error needs to be accounted for. \citet{Berner2017} make the argument that in nonlinear systems, even Gaussian stochastic noise will have an impact on systematic model error. The absence of a fluctuating subgrid-scale has been suggested as one reason for persistent biases across different centers and model versions \citep[][]{Berner2012,Palmer2001} and the omission of this effect might result in compensating model errors. Recently, \citet{Crawford2020} showed that for some surface variables including the random component of tendency corrections from DA was more beneficial than including the deterministic one.

The rise of machine learning in atmospheric science has led to recently renewed interest in tendency adjustments learned from DA systems, as it could lead to state-dependent error parameterizations \citep[][]{DelSole2008}. Online state-dependent bias correction has been previously proposed \citep[][]{DelSole1999,Leith1978,Saha1992}, and implemented in prototype \citep[][]{Brajard2021,Danforth2007} and in full GCMs \citep[][]{Chen2022,Yu2014,Yu2014_v2,WattMeyer2021,bonavita2020machine} which showed success in improving the modeled climate state.

To our knowledge, this is the first study to compare the two DA techniques for use in climate simulations and test the difference in the application of these tendencies in a long-running, state-of-the-art climate model. We set out to answer three main questions with this study:
\begin{enumerate}
    \item To what degree do model-error estimates from nudging tendencies agree with those from EAKF analysis increments?
    \item To what extent does re-inserting DA increments and nudging increments during model runtime reduce climatological model bias of the free-running model?
    \item Will representing subgrid-scale uncertainty in online increment corrections via stochasticity help to improve low-frequency modes of variability without degrading mean state climatological bias?
\end{enumerate}

In the following \textit{Section 2}, we detail the methods and model configurations leveraged for training and implementing the tendency adjustment. Results are presented in \textit{Section 3}. \textit{Section 4} provides a discussion. \textit{Section 5} is a summary and conclusion.

\section{Methods}

\subsection{Analysis Increments}

Data-assimilation increments are leveraged from the Finite Volume Community Atmosphere Model version 6 (CAM6-FV) as a CAM-DART reanalysis was recently produced \citep[][]{Raeder2021} for the period, 2011-2019, referred to henceforth as the training period. \citet{Raeder2021} leveraged an Ensemble Adjustment Kalman Filter (EAKF, Anderson, 2001), which is a state-of-the-art DA system that has been applied to CAM6-FV via the Data Assimilation Research Testbed (DART) \citep[][]{Karspeck2018,Raeder2012}. The EAKF combines an ensemble of short-term forecasts with observations to generate an improved state estimate of the atmosphere at a specific time. By comparing the model simulations to observed data, the EAKF quantifies the error and adjusts the ensemble of model simulations accordingly. This adjustment process reduces uncertainty in the model predictions and brings the ensemble closer to the observed reality.

The reanalysis is the result of an 80-member ensemble of the global atmosphere using CAM6-FV from CESM version 2.1. The archived data represent the actual states of the atmosphere in the training period at horizontal resolutions of $\sim$0.9° latitude and 1.25° longitude at 6-hourly temporal frequency. CAM6-FV has 32 hybrid $\sigma$/pressure levels which are terrain following near the surface and constant pressure near the model top. When producing this reanalysis dataset, the ensemble mean prior and posterior model states were archived. The difference between these states provides the model-tendency adjustments in the 6-hour assimilation window. We refer to the difference of the prior and posterior as the DART model increment and divide by the temporal length of the assimilation window (6 hours) to determine the model-correction tendency (per model time-step). We run so-called CESM “F” compset with prescribed ocean and sea-ice coverage data provided by the AVHRR dataset at a spatial resolution of 0.25°. The temporal resolution is nominally daily but represents an ocean state which is a weighted average of data from several consecutive days \citep[][]{Casey2010}. CAM-FV was run with a standard half-hour time step. Additional external forcing of CAM6-FV includes specified aerosols, greenhouse gases, and volcanic forcing from CESM datasets. These are historical datasets through the end of 2014 and use CMIP6 scenarios from 2015 onwards. Observational data is never assimilated in the top two layers of the model ($\sim$5 hPa - $\sim$2 hPa) as this is known to cause model stability issues. The exact model configuration used in this study for every climate run is permanently archived on GitHub (\url{https://github.com/kdraeder/cesm}), and the data are stored in the National Center for Atmospheric Research’s (NCAR) Research Data Archive.

\subsection{Nudging \& Model Tag}

As a point of comparison to the DART increments, we used a dynamical core-independent nudging scheme that applies a relaxation to observations as a physics tendency controlled via the “nudging” namelist (see Sect. 9.6 of the \href{https://ncar.github.io/CAM/doc/build/html/index.html}{CAM6 user’s guide}). The nudging tendency is applied as a linear relaxation of the form:

\begin{equation}
\frac{\text{d\ x}}{\text{dt}}=-W\ \frac{x-x_{\text{ref}}}{\tau}\nonumber \\
 \end{equation}

Where \(x_{\text{ref}}\) represents the reference meteorology value at the subsequent reference meteorology update step, and \(\tau\) is the relaxation timescale. \(W\) signifies a scalar window function that restricts the spatial extent within which the nudging tendency is implemented. In this case, \(W\) is set to 1 in the full atmosphere and set to 0 in the top 5 layers of the model (approximately 30 hPa - 2 hPa) as nudging near the model top is known to cause stability issues. Additionally, the scheme calculates \(x_{\text{ref}}\) as a linear interpolation between the two nearest reference meteorology values. In computing the update tendencies, the number of dynamical steps taken each day was fixed at 48 (30-minute timesteps), and the tendencies were applied at every timestep using the linear interpolation as described above. The model is nudged towards the ERA-interim reanalysis dataset \citep[][]{Dee2011}, which has a temporal frequency of 6 hours. The reanalysis data was interpolated to the model grid using the bilinear interpolation in the xESMF package. The reanalysis data is interpolated linearly in log pressure coordinates and adjustments are made to account for topographical differences between CAM and the reanalysis product (see Sect. 9.6 of the \href{https://ncar.github.io/CAM/doc/build/html/index.html}{CAM6 user’s guide}).

The nudging can be applied to the model prognostic variables (Zonal [U] and Meridional [V] winds, Temperature [T], specific humidity [Q], and surface pressure [PS]). Through testing multiple nudging configurations, we found that the most effective treatment of model bias is accomplished by adjusting model \(U\) and \(V\) alone when generating the nudging increments (validation was performed over the testing period, not shown). Additionally, multiple relaxation timescales were tested (from 2 hours to 10 hours at 1-hour intervals), and 6 hours was found to be optimal by quantitative examination of the climatological biases (not shown) when generating the nudging increments for the subset of model years 2011-2019. This is convenient as it is analogous to DART’s 6 hour assimilation cycle.

To ensure a fair comparison with DART, the same model tag, lower boundary constraints, and external forcing datasets/configurations were used to generate the nudging increments. The CAM-FV model was again run from January of 2011 through August of 2019 while archiving the instantaneous nudging increments and model state with a temporal frequency of 6 hours. When re-inserting the archived tendencies, the model was run from 1982-2010. This is analogous to splitting data into a testing (1982-2010) and training (2011-2019) dataset so that the increments learned from the model runs in the testing period cannot artificially inflate the model skill or bias the background variability towards observations, once reinserted into the model.

\subsection{Reinserting Adjustment Tendencies}

\begin{figure}[H]
\centering
\includegraphics[width=14cm]{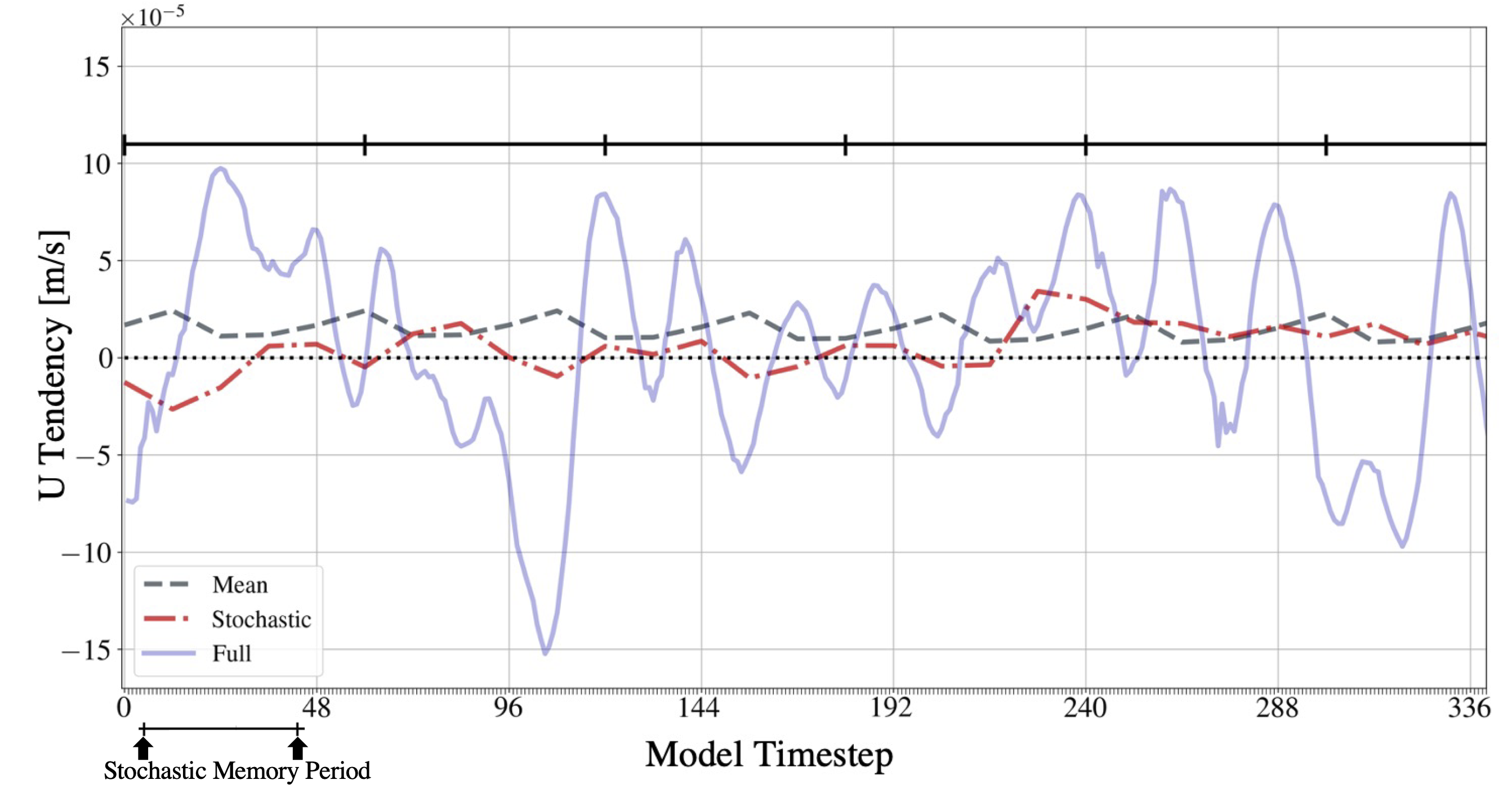}
\captionsetup{}
\caption{Example for the tendency adjustment at a latitude of 15°N and longitude of 140°W for model days Jan 1-14, 1982 (30 minute timestep) in the Nu.S1.M.5 experiment (see Tab. \ref{tab:experiments}). The full model tendency (purple), mean adjustment (dashed gray), and stochastic (dashed red) adjustment are shown. The black ticked line represents the period of system memory for the stochastic increment (see text).}
\label{fig1}
\end{figure}

Fig. \ref{fig1} shows an example of the tendency adjustment (online bias correction) at a single model grid point (15°N, 140°W) over the first 7-days of a model run. The total tendency adjustment added to U and V at every time step is:

\begin{equation} \label{eq:1}
\frac{dx_{t}}{\text{dt}}=f\left(x_{t}\right)+\delta \bar{x}_{t}^{d} + \delta x\sp{\prime d}_{t},\  \\
 \end{equation}

\vspace{12pt}

where $f(x_t)$ is the tendency term of the prognostic model equation and $\delta \bar{x}_{t}^{d} + \delta x\sp{\prime d}_{t},$ is the online bias correction divided into climatological (Fig. \ref{fig1}; grey dash) and stochastic (Fig. \ref{fig1}; red dash) terms, respectively. 

Climatological increment tendencies were archived from both the DART and nudging systems. These are any nonzero long-term averages of the nudging or DART increments, which indicates that these “tendency biases” force the model to drift away from the observational attractor to its own biased climatology. The term bias, in this text refers to the time mean differences between the model forecasts and observations (reanalysis product). The climatological increment (from either DART or nudging) is calculated by taking a centered, 31-day climatology of the system increments while respecting the model time of day (at 6-hour increments). The functional form of the mean increment adjustment is thus, $\delta \bar{x}_{t}^{d} =\frac{\alpha}{{N_{s}}^{d}}\ \sum_{k=1}^{{N_{s}}^{d}}{\delta{x_{k}}^{d}}$, where the mean increment ($\delta \bar{x}_{t}^{d}$) is conditioned on the model day $t$ and time of day ($d$, at 6-hour interval \(\in[0,6,12,18]\ \text{hrs}\)). \({N_{s}}^{d}\) is the total number of increments in a 31-day window spanning the 10-year training period at each model time of day (310 new samples to select from each time-step). This accounts for system increment evolution in both the seasonal and diurnal cycle (which were found to be significant). 
 
Following \citet{batte2016_new} and \citet{Crawford2020}, a library of anomaly increments ($\delta x\sp{\prime d}_{t}$), was archived, and provided as additional tendency terms to the model through stochastic selection (Fig. \ref{fig1}, red dash). The anomaly tendencies were randomly selected from the same centered 31-day period, while again respecting the model time of day. The total tendency inserted follows the form of equation \ref{eq.2}. The first term is the model mean increment, and the second term is a stochastically selected increment anomaly. The anomaly term is read forward in time for \(p\) timesteps where \(p\) is incremented from 0 to 120, and then \(\delta{x_{r_{i+p}}}^{d}\) is randomly sampled again. The anomaly increment library integrates forward 120 timesteps (30 hours) to retain some temporal autocorrelation of the stochastic increments. \(\alpha\) and \(b\) are scaling terms to control the magnitude of the stochastic and mean portion of the tendency, respectively. Here, we set \(\alpha\) = 1 and \(b\) = 0.5 (after multiple rounds of validation [by miniizing the climatological bias of surface temperature and precipitation] on the training period [2010-2019] for optimal results; not shown).

\begin{equation} \label{eq.2}
\delta x_{t} = \delta \bar{x}_{t}^{d} + \delta x\sp{\prime d}_{t} =\frac{\alpha}{{N_{s}}^{d}}\ \sum_{k=1}^{{N_{s}}^{d}}{\delta{x_{k}}^{d}}+b\left(\left[\delta{x_{r_{i}+p}}^{d}\right]-\left[\frac{\alpha}{{N_{s}}^{d}}\ \sum_{k=1}^{{N_{s}}^{d}}{\delta{x_{k}}^{d}}\right]\right)\ \\
 \end{equation}

We refer to the first term of equation \ref{eq.2} as a mean increment tendency adjustment (MITA, $\delta \bar{x}_{t}^{d}$) and the second half as a stochastic increment tendency adjustment (SITA, $ \delta x\sp{\prime d}_{t}$), and name our experiments following the convention: Adjustment System\_MITA*$\alpha$\_SITA*$b$. For example, a tendency adjustment derived from the DART system which applies a mean adjustment scaled to 1 and a stochastic adjustment scaled with 0.5 would be named: DArt\_M1\_S.5 (nudging would be Nu\_M1\_S.5). The experiments analyzed in this study are shown in Table \ref{tab:experiments}.

\begin{longtable}{@{}llp{3cm}p{3cm}l@{}}
\label{tab:experiments}\\
\toprule
\textbf{Expt. No.} & \textbf{Expt. Name} & \textbf{Description} &
\textbf{Model} & \textbf{Lower Boundary} \\
\midrule
\endhead 
1 & CNTRL & 29 yr Control run spanning 1982-2010 & AGCM & Prescribed SST \& Sea-Ice \\
2 & DArt.M1.S.0 & 29 yr MITA run spanning 1982-2010 & AGCM with the
addition of MITA{[}1{]} with tendencies derived from DART & Prescribed SST \& Sea-Ice \\
3 & Nu.M1.S.0 & 29 yr MITA run spanning 1982-2010 & AGCM with the
addition of MITA{[}1{]} with tendencies derived from the nudging & Prescribed SST \& Sea-Ice \\
4 & DArt.M1.S.5 & 29 yr MITA/SITA run spanning 1982-2010 & AGCM with the
addition of MITA{[}1{]} \& SITA{[}.5{]} with tendencies derived from
DART & Prescribed SST \& Sea-Ice \\
5 & Nu.M1.S.5 & 29 yr MITA/SITA run spanning 1982-2010 & AGCM with the
addition of MITA{[}1{]} \& SITA{[}.5{]} with tendencies derived from
nudging & Prescribed SST \& Sea-Ice \\
\bottomrule
\caption{Description of the experiments analyzed in this study}
\end{longtable}

\begin{figure}[H]
\centering
\includegraphics[width=15cm]{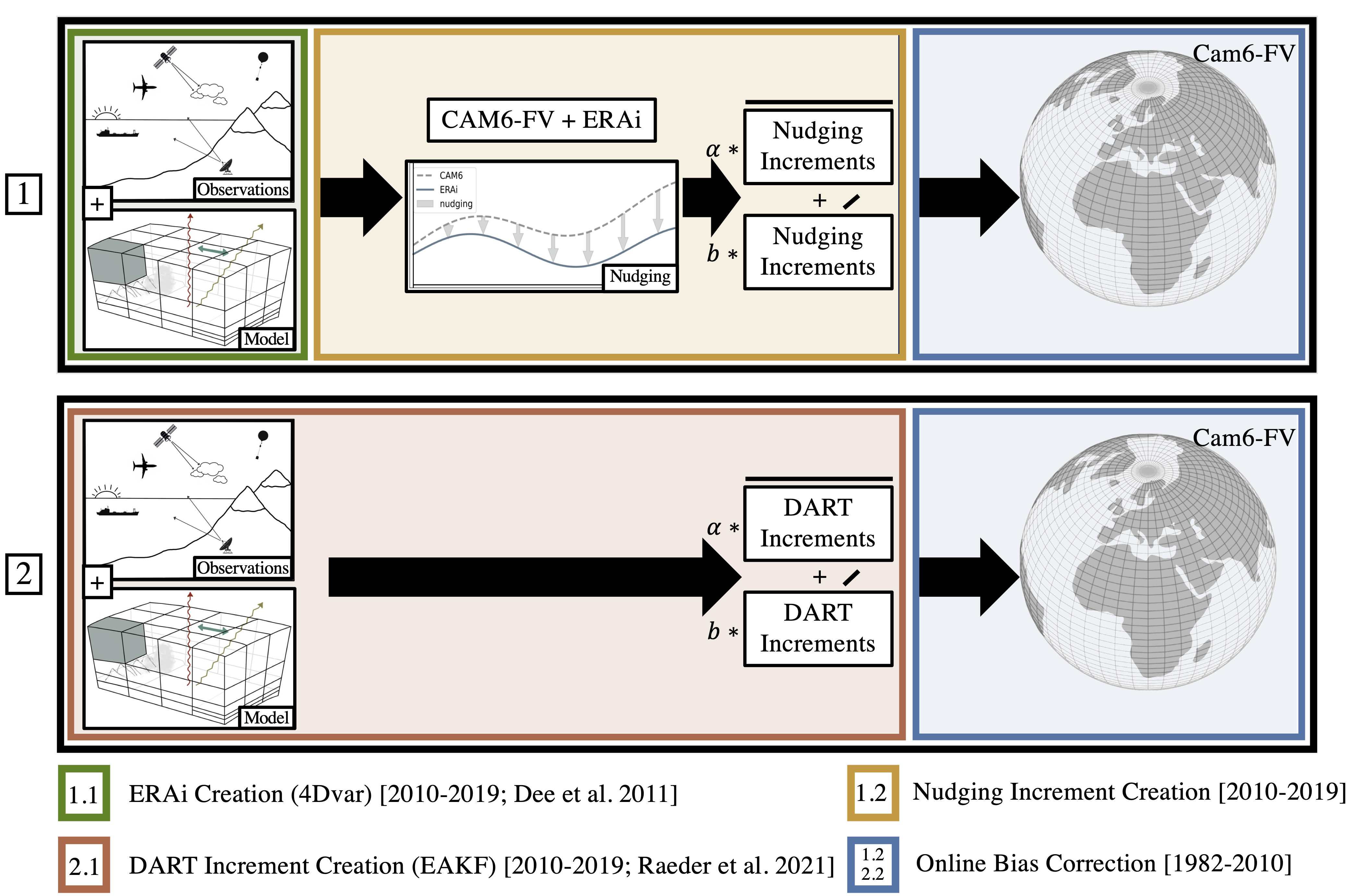}
\captionsetup{}
\caption{MITA/SITA workflow diagram showing the generation (red, yellow, and green box) and online implementation (blue boxes) of nudging (1) and DART (2) DA increments. The green box (1.1) shows the generation of the ERA-interim reanalysis data set (Dee et al., 2011) [model years 2011-2019]. The yellow box (1.2) shows the creation of the nudging increments via 1 run of CAM6-FV [model years 2011-2019]. The red box (2.1) shows the generation of the CAM6-FV+DART reanalysis [80 ensemble members, model years 2011-2019]. The blue box (1.2/2.2) shows the online implementation in CAM6 [model years 1982-2010].}
\label{fig2}
\end{figure}

Fig. \ref{fig2} provides a comprehensive visualization of this study's workflow, which commences with two simultaneous processes for generating adjustment increments during the training years of 2011-2019: (1) nudging to ERAi (shown by the green and yellow boxes) and (2) the DART DA system (indicated by the red box). Following their generation, these increments are averaged to form a climatology of corrective increment forcing. This climatology is subsequently employed to generate a library of anomaly increments, which are used for stochastic selection. The resulting terms are multiplied by specific scalars (\(\alpha\), \(b\)) before being integrated into the online model's runtime during the testing years of 1982-2011 (depicted by the blue box). The workflow concludes with an evaluation of model bias within the online runs, thereby shedding light on the effectiveness of the implemented adjustments.

\subsection{Streamfunction Tendency}
To clarify the relative roles of various dynamical processing in influencing the evolution of tropospheric low-frequency circulation, a decomposed streamfunction tendency equation \citep[][]{Cai1994} is diagnosed as in \citep[][]{michel2011link,riviere2015dynamics,tan2017role}:

\begin{equation}\label{eq.3}
\frac{\partial\psi^{L}}{\partial t}=\ \sum_{i=1}^{5}\xi_{i}\ +\ R\ 
 \end{equation}
where:
\begin{gather*}
\xi_{0} = \nabla^{-2}(-\zeta^{L}) = \frac{\partial\psi^{L}}{\partial t} \\
\xi_{1} = -\nabla^{-2}{\{\overline{\mathbf{V}}\cdot\nabla\zeta^{L}\ +\ \ \zeta^{L}\nabla\cdot\overline{\mathbf{V}} \}}^{L} \\
\xi_{2} = -\nabla^{-2}{\{\mathbf{V}^{L}\cdot\nabla\overline{(f+\zeta)} + \overline{(f+\zeta)}\nabla\cdot\mathbf{V}^{L}\}}^{L} \\
\xi_{3} = \nabla^{-2}{(-\mathbf{V}_{r}^{H}\cdot\nabla\zeta^{H})}^{L} + \nabla^{-2}\{-\nabla{\cdot(-\mathbf{V}_{d}^{H}\zeta^{H})\}}^{L} \\
\xi_{4} = \nabla^{-2}{(-\mathbf{V}_{r}^{L}\cdot\nabla\zeta^{L})}^{L} + \nabla^{-2}\{-\nabla{\cdot(-\mathbf{V}_{d}^{L}\zeta^{L})\}}^{L} \\
\xi_{5} = \nabla^{-2}{(-\mathbf{V}_{r}^{L}\cdot\nabla\zeta^{H})}^{L} + \nabla^{-2}\{-\nabla{\cdot(-\mathbf{V}_{d}^{L}\zeta^{H})\}}^{L} + \nabla^{-2}{(-\mathbf{V}_{r}^{H}\cdot\nabla\zeta^{L})}^{L} + \nabla^{-2}\{-\nabla{\cdot(-\mathbf{V}_{d}^{H}\zeta^{L})\}}^{L}
\end{gather*}

Here, \(\psi^L\) is the streamfunction, \(\zeta\) is the relative vorticity, \(V\) is the horizontal wind vector, and \(f\) is the Coriolis parameter. \(\nabla^{-2}\) represents an inverse Laplacian operator. The term \(R\) indicates the residual term induced by processes such as dissipation, external forcing, and two neglected terms [vertical advection and twisting terms \citep[see,][]{Feldstein1998}]. The subscripts \(r\) and \(d\) represent the rotational and divergent components of the wind. The superscripts \(H\) and \(L\) indicate the high- and low-frequency components that are divided at a period of 10-days (we then remove the climatology to form the low-frequency anomalous wind). All frequency filtering is done using Fast-Fourier transform, and the climatological wind is the centered 90-day average of the entire model simulation (29-year run). The low-frequency tendency of the streamfunction (\(\xi_0\)) is calculated as centered differences, except for the first (last) time step where a forward (backward) finite-difference scheme is used, this term is calculated as a point of comparison for the full decomposition. The calculations are performed on daily files, and thus a timestep of 24hrs is used, and the data is interpolated to a 1.5° regular grid when the tendency is performed. The physical interpretations of the processes associated with \(\xi_1 - \xi_5\) are (i) the low-frequency eddy vorticity advection by the climatological wind and its divergence term, (ii) the climatological absolute vorticity advection by the low-frequency wind anomaly and its divergence term, (iii) the interaction among all the low-frequency transient eddies, (iv) the self-interaction among the high-frequency anomalies, and (v) the cross-frequency interaction between high- and low-frequency anomalies.

As discussed in \citet{Cai1994}, there is some arbitrariness in the tendency decomposition, which comes fully or partly from the separation of the high- and low-frequency transients and the separation of rotational wind from irrotational wind. A full step-by-step method of this calculation is provided in the supplemental material, and a Jupyter notebook to perform the calculation is provided in the paper's public GitHub repository, which performs this decomposition on observations from daily CAM6-FV model output. The first two terms in equation \ref{eq.3} (\(\xi_1\) \& \(\xi_2\)), are designated as linear terms, and the subsequent three terms (\(\xi_3 - \xi_5\)) are designated as nonlinear terms. All gradients and the inverse Laplacian calculations are done with spherical harmonics in the windspharm package (Dawson, 2016).

The physical processes that are responsible for the growth and decay of any persistent low-frequency pattern (i.e., the Pacific North American (PNA) pattern, or the North Atlantic Oscillation (NAO)) can be clarified explicitly by examining each term on the RHS of eq. 3. This can be summarized by projecting each term onto the composite anomalous streamfunction pattern on a specified day (M). The projection for each \(\xi_i\) can be written as \citep[][]{Feldstein2003,Xu2020,Shen2023}:

\begin{equation}
P_{i} = \sum_{\lambda,\theta} \frac{\xi_{i}(\lambda,\theta)\psi_{M}(\lambda,\theta)\cos \theta}{\psi_{M}^{2}\cos \theta} \quad (4) \nonumber \\
\end{equation}

Where \(\psi_M\) is the streamfunction pattern on day M; \(\lambda\) and \(\theta\) are the longitude and latitude, respectively. The projection is then performed over a region, which is specified in each figure, and shows the time rate of change of \(\psi^L\) towards day M.

\subsection{Significance}

Significance is assessed via bootstrapping.  Bootstrapping is a statistical technique used to assess significance by estimating the sampling distribution of a statistic without making assumptions about the data distribution. In this approach, a designated sample size (N, usually $\sim$ 500-10000) is drawn from the original data with replacement to create a bootstrap sample distribution. The statistic of interest (e.g., climatological/seasonal mean bias) was calculated on this sample, and the process is repeated N times to create a distribution of the statistic. This distribution was then used to estimate pseudo confidence intervals.

We used a statistically strict criterium to assess significant difference. When comparing two simulation or observation statistics, as in \citet{deser2017northern}, two distribution samples are created, and their overlap at the 5th/95th percentile is examined to determine statistical significance. If there is an overlap, we conclude that there is no statistical difference between the two samples.

Various methods of model validation are performed in the results section, and we refer the reader to the appendix for a detailed explanation of the calculated empirical orthogonal functions (EOF), blocking index calculation, and global bias calculation used in this study.

\section{Results}

\begin{figure}[H]
\centering
\includegraphics[width=12cm]{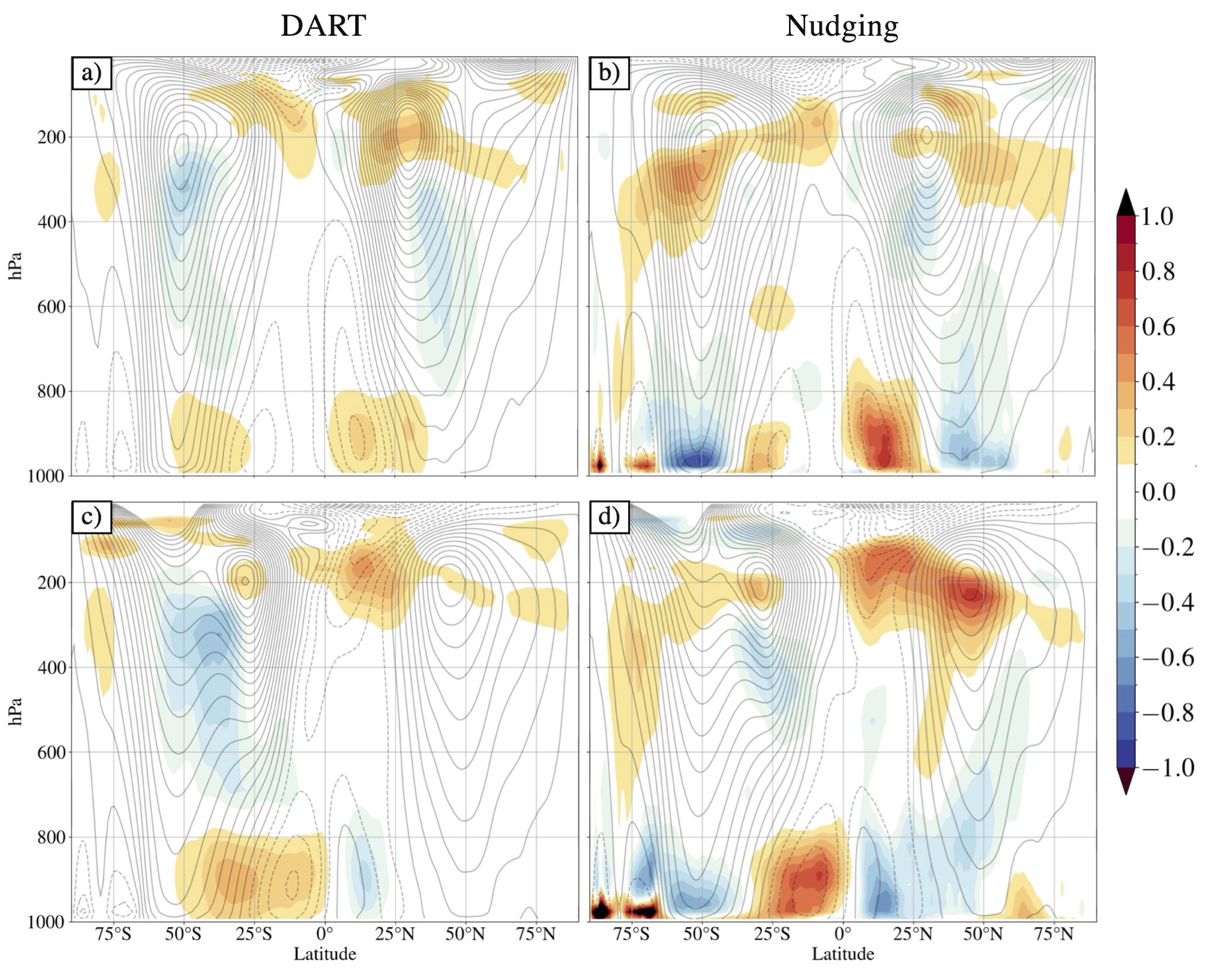}
\captionsetup{}
\caption{The zonal-mean \(u\) DA increments (\(m\,s^{-1}\,d^{-1}\)) in DJF (a, c) and JJA (c, d) for the DART (Column I) and nudging system (Column II). Contours show the \(u\) wind climatology [2 \(m\,s^{-1}\) intervals, negative is dashed]. All fields are averaged over the period 1982-2010.
}
\label{fig3}
\end{figure}

Figure \ref{fig3} shows a vertical cross section of the zonal-mean U wind increment derived from the DART (column 1) and nudging (column 2) data in DJF (a, d) and JJA (b, e). Systematic increments occur at every vertical level of the model. In the lower-level tropics, a clear seasonal cycle is present. Consistent with \citet{Simpson2018}, surface increments act to reduce the easterly flow at low latitudes where the climatological (Fig. \ref{fig3}; contours) low-level easterly flow is the strongest across the trade-wind region. The increment magnitude scales with lower-level wind strength, leading to a larger increment present in the winter-time hemisphere in both the DA and nudging data.

The sign of the increments differs over the low-level Southern Ocean $\sim$ [40°S, 60°S, 975 hPa], where the nudging increments act to dampen the anomalously strong westerly flow. In this region, the nudging tendencies also exhibit a strong seasonal cycle, which again scales with the strength of the climatological westerly flow. This signal is not present in the DA system, but this is potentially a consequence of the sparsity of observations assimilated in this region \citep[][]{Raeder2021}. At the upper levels ($\sim$ 200 hPa), both datasets show a systematic strengthening of the Northern Hemisphere jet. In the mid-troposphere Southern Hemisphere, opposite signed increments are present that are most clearly seen in DJF. Supplemental Fig. S1 shows the same figure but for the meridional winds. In both seasons, the increments act to strengthen the upper-tropospheric southerlies and the lower-tropospheric northerlies, which generally indicates that the model has too weak Hadley cell overturning.

\begin{figure}[H]
\centering
\includegraphics[width=10cm]{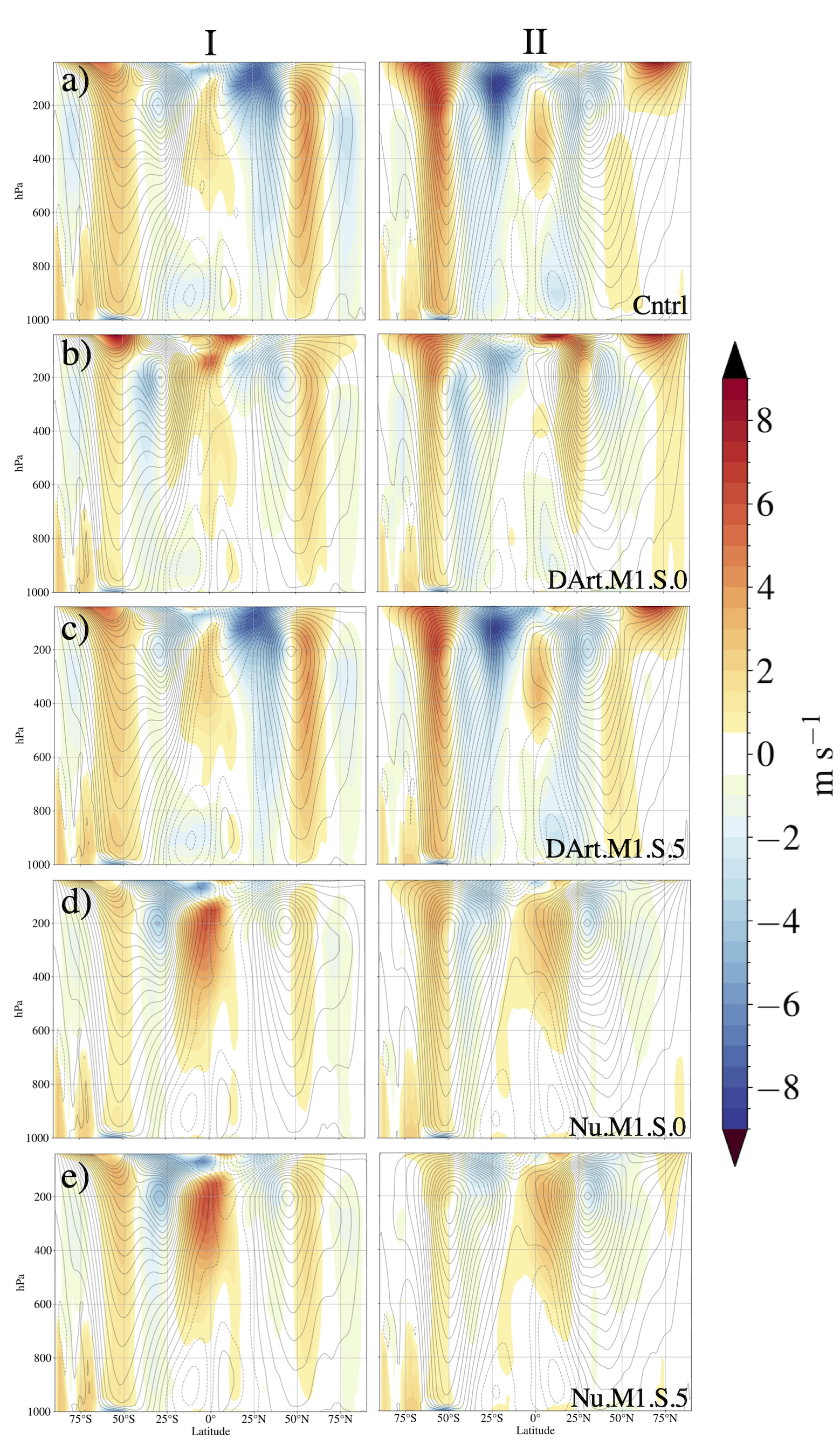}
\captionsetup{}
\caption{The zonal-mean \(u\) bias (\(m\,s^{-1}\)) in JJA (column I) and DJF (column 2) for each experiment listed in Table \ref{tab:experiments}. Contours show the \(u\) wind climatology [2 \(m\,s^{-1}\) intervals, negative is dashed]. All fields are averaged over the period 1982-2010.
}
\label{fig4}
\end{figure}

The impact of the tendency bias correction on the zonal-mean climatological biases is explored in Figure \ref{fig4} for JJA (column I) and DJF (column II). Here, the bias is shown as the difference between the seasonal model runs and the ERA-interim seasonal mean (Experiment - ERA, therefore positive (negative) values show a positive (negative) bias). The zonal-wind biases in Cntrl are characteristic of a poleward shift of the midlatitude jets in the summer hemisphere (see the dipole $\sim$200 hPa. Fig. \ref{fig4}a). Marked positive impacts of the online bias correction scheme can be seen in the full atmosphere for the DArt.M1.S0, Nu.M1.S0, and Nu.M1.S.5 runs. At upper-levels, the jet bias is improved in both seasons and hemispheres. At lower-levels the nudging MITA (Nu.M1.S0, and Nu.M1.S.5) particularly improves the U wind biases, entirely eradicating them in some locations. An area that appears to be degraded is in the mid-to-upper-troposphere in the nudging MITA/SITA runs (Fig. \ref{fig4}d \& \ref{fig4}e, 20°S-20°N); however, this an artifact of the zonal mean. The CNTRL, DArt.M1.S0, and DArt.M1.S.5 simulations have compensating biases in the tropics which are out of phase and thus cancel and present as a small bias in the zonal average (refer to Table S2 and Fig. \ref{fig5}, to see the bias in the tropics and a planar view of the bias), when in fact the CNTRL, DArt.M1.S0, and DArt.M1.S.5 biases are larger than the biases in the Nu.M1.S0, and Nu.M1.S.5 systems. 

\subsection{Seasonal Global Mean Bias}

\begin{figure}[bt]
\centering
\includegraphics[width=15cm]{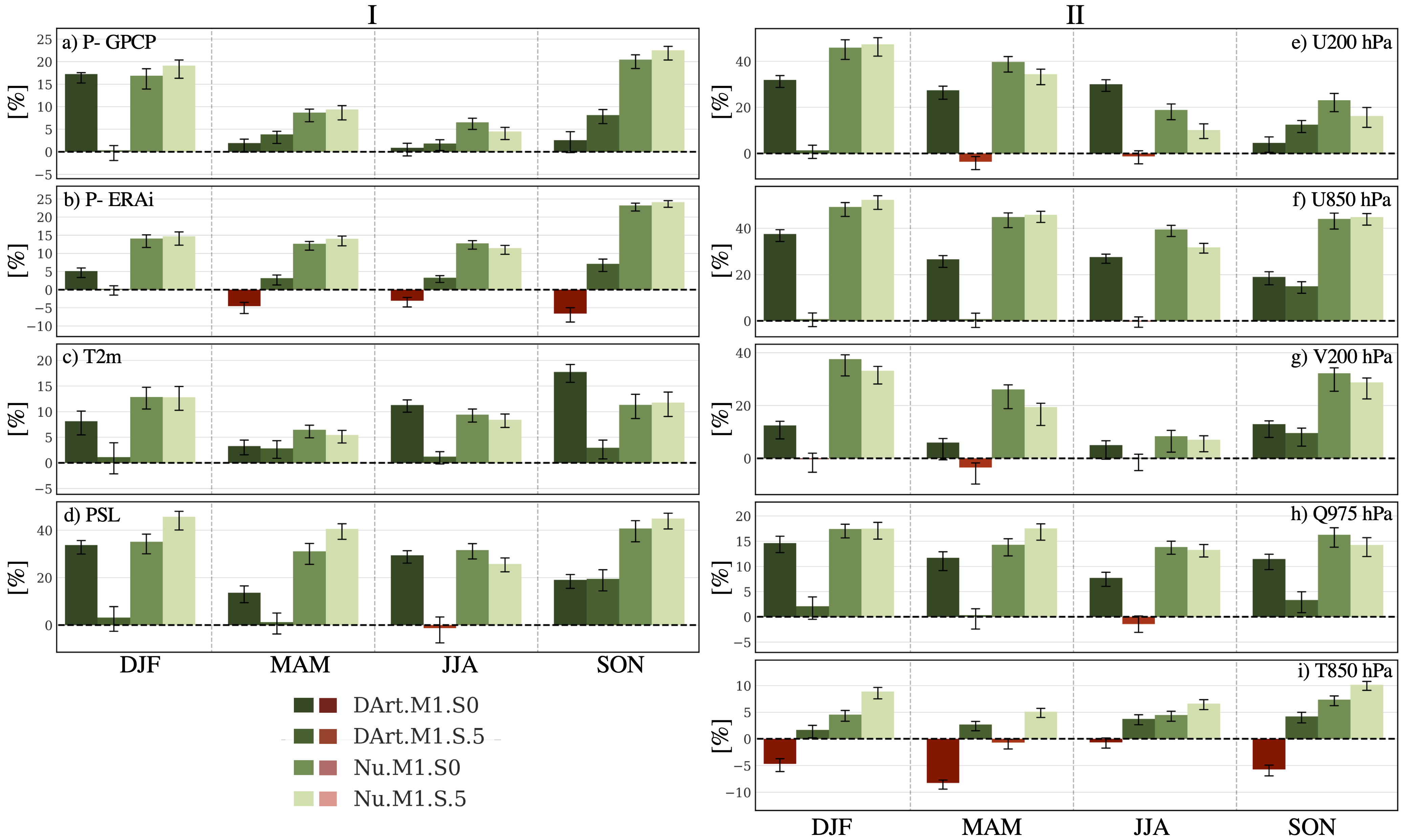}
\captionsetup{}
\caption{Global percent improvement over the Cntrl run in all seasons, for variables Precipitation (a; ERAi-observations, b; NOAA GPCP observations), T2m (c), PSL (d), U200 hPa (e), U850 hPa (f), V200 hPa (g), Q975 hPa d(h), T850 hPa (i). Error bars show the 5th and 95th percentile from the synthetic bootstrap distribution. All fields are averaged over the period 1982-2010. Each model is shown in different color, positive results are in green, negative results are in red (see figure key).
}
\label{fig5}
\end{figure}

We examine the percent improvement over the control run in global mean bias (Fig. \ref{fig5}) for surface variables (precipitation [both P-GPCP and P-ERAi], surface temperature [T2m], and sea-level pressure [PSL]; Fig. \ref{fig5}, column I) and variable bias at multiple vertical levels for zonal wind, meridional wind, specific humidity, and temperature (Fig. \ref{fig5}, column II). Two observational products were used to estimate precipitation bias - the NOAA global precipitation climatology project monthly precipitation analysis (P-GPCP) and the ERA-interim derived precipitation (P-ERAi) - since they have different precipitation climatologies, especially over the ocean \citep[][]{Nogueira2020}.

In general, the online bias correction demonstrates a positive impact on global biases across all examined variables, resulting in significant improvements in most seasons. Instances of model degradation are rare. In the following text, we highlight specific findings to underscore key results. For a comprehensive assessment, we provide a detailed analysis of model bias (as definied in the appendix) for multiple variables, including prognostic, diagnostic and transients in supplemental Tables S1-S3. These tables also offer a breakdown of bias information based on regional divisions, specifically the tropics [25°S-25°N] and extratropics [90°S-25°S; 25°N-90°N], as well as distinctions between land and ocean regions.

Through a comprehensive examination of the Nu.M1.S0 and Nu.M1.S.5 (lighter greens and reds, Fig \ref{fig5}.) experiments, it is evident that the online bias correction yields significant improvements across all seasons for both the deterministic and stochastic approaches. Particularly noteworthy are the substantial enhancements observed during the cool season in the Northern Hemisphere. The prognostic variables show a notable reduction in bias, ranging from 10\% to 40\%. Moreover, diagnostic variables show improvements of approximately 10\% to 20\%. It is important to note that adjustments were made solely to zonal and meridional winds, while the enhancements in non-dynamic variables result from dynamic interactions with the corrected wind field.

\begin{figure}[H]
\centering
\includegraphics[width=12cm]{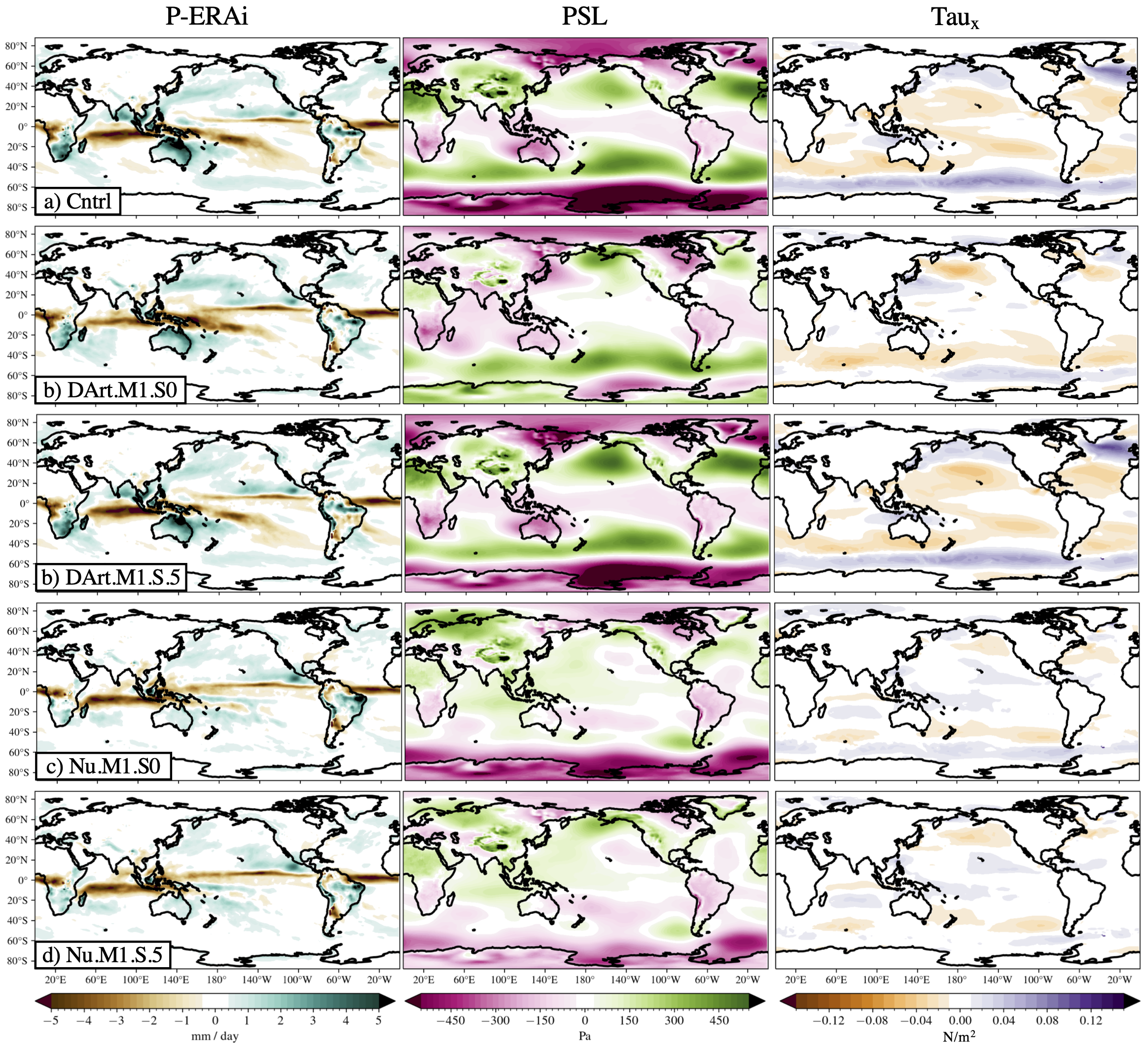}
\captionsetup{}
\caption{DJF bias in Precipitation (mm/day, left), PSL (Pa, middle), and surface wind stress (N/m$^2$, right). Bias is oriented as field – observations (positive (negative) numbers are biased high (low)), for the five model configurations (a-d). All fields are averaged over the period 1982-2010.
).
}
\label{fig6}
\end{figure}

Figure \ref{fig6} illustrates the spatial maps of P-ERAi, PSL, and zonal surface wind stress (TAUx, over ocean) biases in DJF, JJA is shown in supplemental Figure S3. In the best case showing substantial improvement in precipitation biases, with a 21\% improvement over land and a 22.8\% improvement over the total extratropics when compared with the CNTRL simulation. Notably, precipitation in the midlatitude Pacific storm track exhibits considerable enhancement, likely indicating improved atmospheric river transport \citep[][]{Chapman2019,Zhu1998}. The adjustments made to sea level pressure (PSL) yield significant improvements, particularly with a 45.7\% improvement globally in DJF, a 25.7\% improvement in JJA, and an annual improvement of 50\% (see Table S1). The biases in the storm-track regions are significantly reduced, and the bias off the coast of Antarctica shows considerable improvement. Zonal surface wind stress (TAUx) also improves, especially in the Southern Ocean. The nudging increments lead to a 52.9\% improvement in DJF and a 28.9\% improvement in JJA in the best experiments.

\begin{figure}[H]
\centering
\includegraphics[width=14cm]{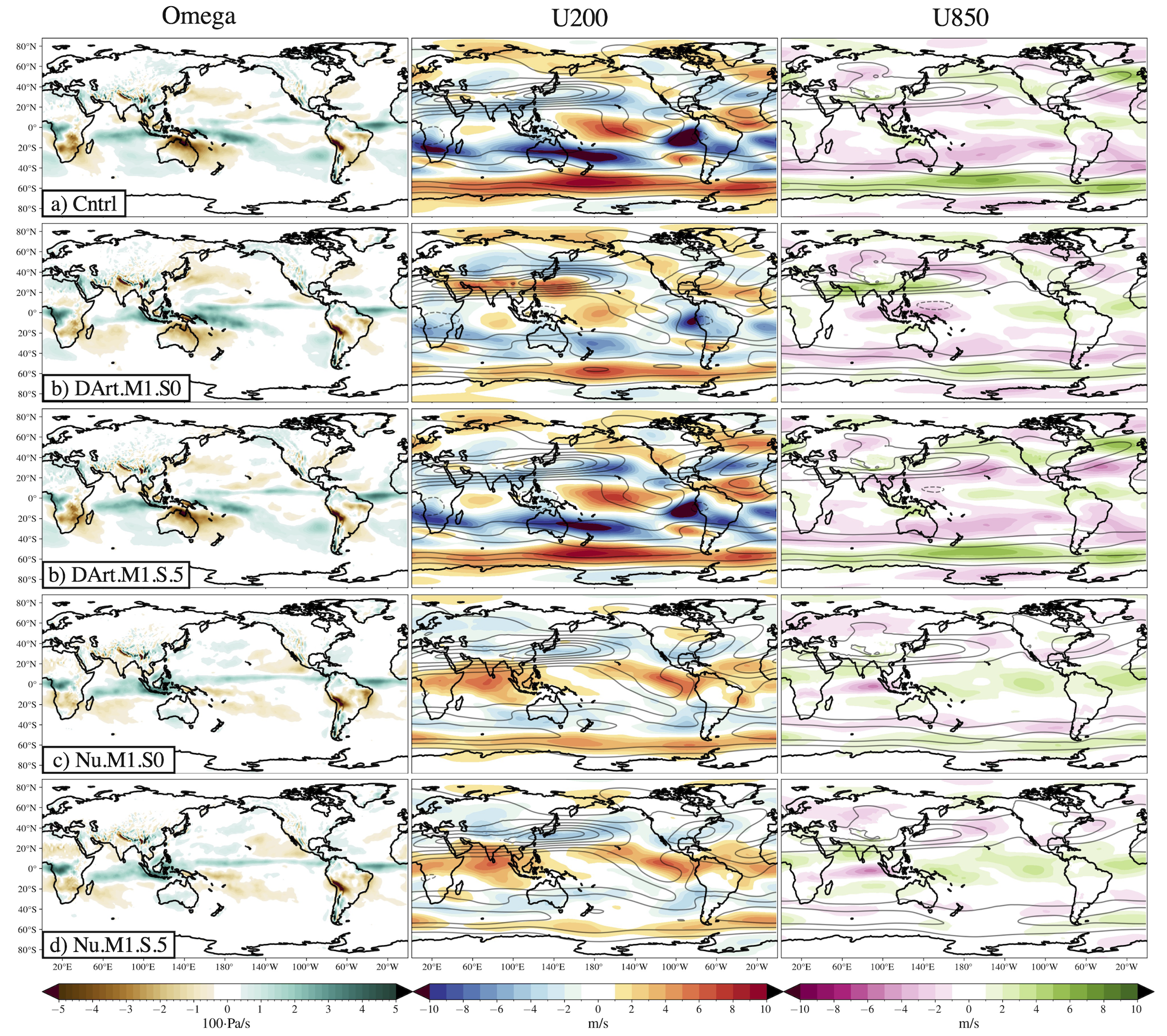}
\captionsetup{}
\caption{DJF Bias in vertical velocity (Pa s\textsuperscript{-1}, left), Zonal 200 hPa wind (m s\textsuperscript{-1}, middle), and Zonal 850 hPa wind (m s\textsuperscript{-1}; right). Bias is oriented as field – observations (positive (negative) numbers are biased high (low)), for the five model configurations (a-d). Contours (middle, right) show the DJF climatological wind field [m s\textsuperscript{-1}; 10 m s\textsuperscript{-1} contour interval (excluding 0), negative indicated by dashed line]. All fields are averaged over the period 1982-2010.
).
}
\label{fig7}
\end{figure}

Figure \ref{fig7} illustrates the significant improvement in winds and dynamic variables resulting from the inclusion of DA and nudging increments. Specifically, W850 (vertical velocity of wind at 850 hPa), U200, and U850 in DJF are shown, with additional information provided in the supplemental material for JJA (Fig. S4). Notably, the nudging increments have a pronounced impact on vertical velocity in the South Pacific convergence zone (SPCZ), influencing tropical precipitation. In JJA (Fig. S4), the nudging schemes lead to a reduction in bias and a more accurate precipitation climatology over the maritime continent. The representation of the summer hemisphere jet exhibits improvements in various model configurations, except for DArt.M1.S.5. Lower-level winds (U850) also show substantial enhancements, particularly over the Southern Ocean and North Atlantic. 
Overall, the inclusion of DA and nudging tendencies results in considerable improvements in global climatology for both DJF and JJA, with the dominant improvements occurring in the summertime jet region and the Southern Ocean.

\subsection{Impact on Transients}

\begin{figure}[H]
\centering
\includegraphics[width=12cm]{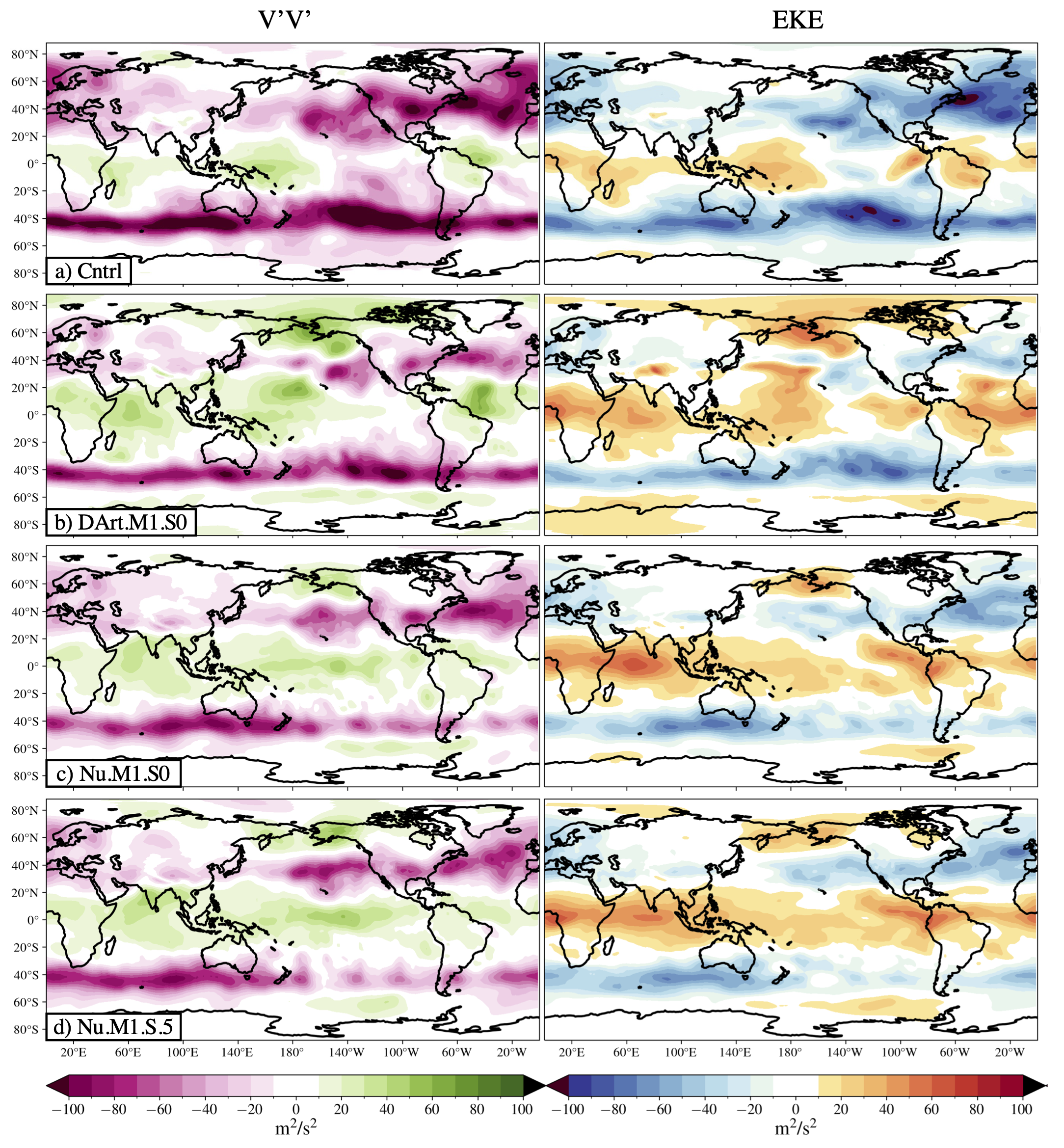}
\captionsetup{}
\caption{DJF Bias in 200 hPa Square of transient component of the meridional wind ($m^2 s^{-2}$; left), 200 hPa transient kinetic energy ($m^2 s^{-2}$, middle), and 850 hPa moisture flux by transients ($g kg^{-1} m s^{-1}$; right). Bias is oriented as field – observations, for the four model configurations (a-d). All fields are averaged over the period 1982-2010.
).
}
\label{fig8}
\end{figure}

Next, we turn our attention to the bias in the transients in DJF (Fig. \ref{fig8}), JJA is shown in the supplemental material (Fig. 5S). These are based on daily data (with the monthly means removed) and we examined the time-mean 250 hPa meridional wind variability $(\overline{v'^2})$, a measure of storm track activity \citep[][]{Chang2002}, and the 200 hPa transient kinetic energy $(EKE, \overline{v'^2+u'^2})$, a measure of mesoscale transient wave activity. No results for DArt.M1.S.5 are shown, because they did not show improvements over the Cntrl run (though little-to-no significant model degradation was observed). The climatological biases in the three shown quantities are apparent, and there are substantial corrections, especially in the summer hemisphere. There are negative biases $(\overline{v'^2})$ which indicate a weak storm track, largely the corrections coincide with the correction to the mean state of zonal wind (Fig. \ref{fig4}), in which the shifted jet location was corrected. For $(\overline{v'^2})$ in DJF the global improvement is 26.5\%, 33.3\% and 31.8\% for the DArt.M1.S0, Nu.M1.S0, and Nu.M1.S.5 simulations, respectively. JJA shows similar significant improvements to $(\overline{v'^2})$ (Table S3). The transient kinetic energy is shown to be too weak across the extratropical region. The MITA/SITA corrections act to improve the EKE across the extratropical regions; however, they are a detriment to the tropics. Globally, there is still a significant improvement in both DJF and JJA, despite the compensating bias arising across the tropical region.

\subsection{Low Frequency Variance Modes}

In the previous section, we showed that the first-order statistics of the climatological state are improved across all examined global metrics by adding MITA/SITA tendency bias corrections. However, any assessment of the health of a climate model must also consider the accurate representation of natural climate variability and low-frequency climate modes \citep[e.g.,][]{Phillips2014}. The dynamical mechanisms associated with intraseasonal variability are complex and act across myriad timescales, leading to societally influential downstream weather \citep[e.g.,][]{Branstator1992,Simmons1983,wallace1981teleconnections}. The mechanisms for the growth and maintenance of these patterns can be compared in the models’ space with observations to assess model health and real-world representation \citep[e.g.,][]{Feldstein1998}. Additionally, the low-frequency variability is highly attenuated by the climatological background state of the model, with many mechanisms having been proposed to maintain and grow this variability. These climatological/transient interplay mechanisms include: growth of the low-frequency anomaly due to instabilities associated with the zonally asymmetric midlatitude jet \citep[e.g.,][]{Branstator1990,Branstator1992,Frederiksen1983,Simmons1983}, changes in quasi-stationary eddies associated with fluctuations in the zonal-mean flow \citep[e.g.,][]{Branstator1984,Kang1990}, forcing from tropical heating or orography \citep[e.g.,][]{hoskins1981steady,sardeshmukh1988generation}, and the vorticity flux from high-frequency eddies  \citep[e.g.,][]{Branstator1992,egger1983theory,Lau1988,Ting1993}.

Here, we explore the PNA pattern \citep[][]{wallace1981teleconnections} and the NAO \citep[][]{Barnston1987}, two of the Northern Hemisphere’s most impactful and dominant modes. Since these patterns are most pronounced and influential in DJF, we narrow the discussion in this manuscript to the Northern Hemisphere winter, as this was the time with the highest biases present in all experiments and the wintertime surface climate variability is more strongly governed by the variability in the atmospheric circulation compared to summer variability \citep[][]{Walker1924}.

\begin{figure}[H]
\centering
\includegraphics[width=14cm]{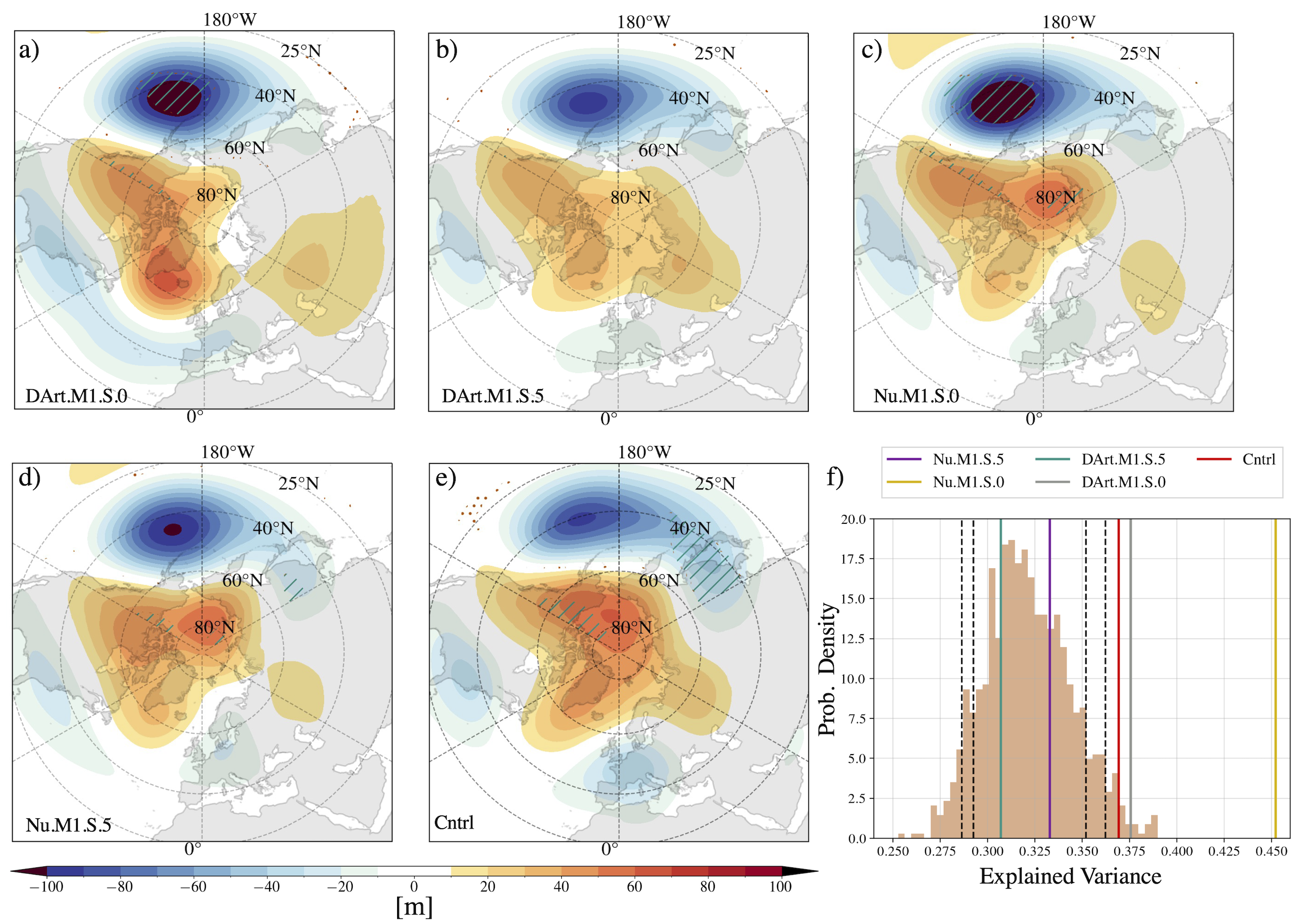}
\captionsetup{}
\caption{DJF 500 mb geopotential height leading mode of variability over the region [PNA, 20-85°N, 120°E-120°W], for every experiment configuration (a-e). Hatching (stippling) shows where the variance in the observations is higher (lower) than the model using the 5th and 95th percentile of the bootstrapped synthetic observations as a bias threshold. Where observational spread is determined by bootstrap (see methods). Also, the explained variance in each model experiment (vertical line), and the bootstrapped spread of explained variance in the observations (histogram) (f), grey vertical dashed lines show the 5th, 10th, 90th, and 95th percentile of bootstrapped observational variance explained.
}
\label{fig9}
\end{figure}

The PNA is the leading mode of low-frequency variability in the Northern Hemisphere Pacific region \citep[][]{Franzke2005,Franzke2011}. It is shown for DJF in Figure \ref{fig9}, where stippled and hashed regions denote a low and high variance bias, respectively. In its positive phase, the PNA is characterized by a deepened Aleutian low, an increased Canadian High, and a deepened Florida Low pattern which extends into the Atlantic \citep[][]{wallace1981teleconnections} (Wallace \& Gutzler, 1981). The Aleutian low limb of the PNA, in particular, is responsible for downstream effects of precipitation and temperature anomalies across the much of North America \citep[e.g.,][]{Carrera2004,coleman2003ohio,liu2017pacific,notaro2006model,chapman2021monthly}. The Cntrl run shows a good representation of the pattern across the North Pacific, with an Aleutian low pattern without bias. However, the explained variance is much higher than in ERAi (Fig. \ref{fig9}f). When adding MITA increments, DArt.M1.S0 and Nu.M1.S0 the Aleutian becomes too deep. Despite having a background state that is relatively unbiased compared to the control run, the leading mode of variability is deepened. Additionally, in DArt.M1.S0, Nu.M1.S0, and Cntrl, the second leading mode of variability in this region (the North Pacific Oscillation) is too weak (not shown) as the PNA dominates the North Pacific variance space. However, the addition of stochastic increments (DArt.M1.S.5 and Nu.M1.S.5) acts to remove variance in the AL. The two stochastic simulations show Aleutian low patterns that are unbiased and explained variances that are not significantly different from the variance seen in observations.

\begin{figure}[H]
\centering
\includegraphics[width=14cm]{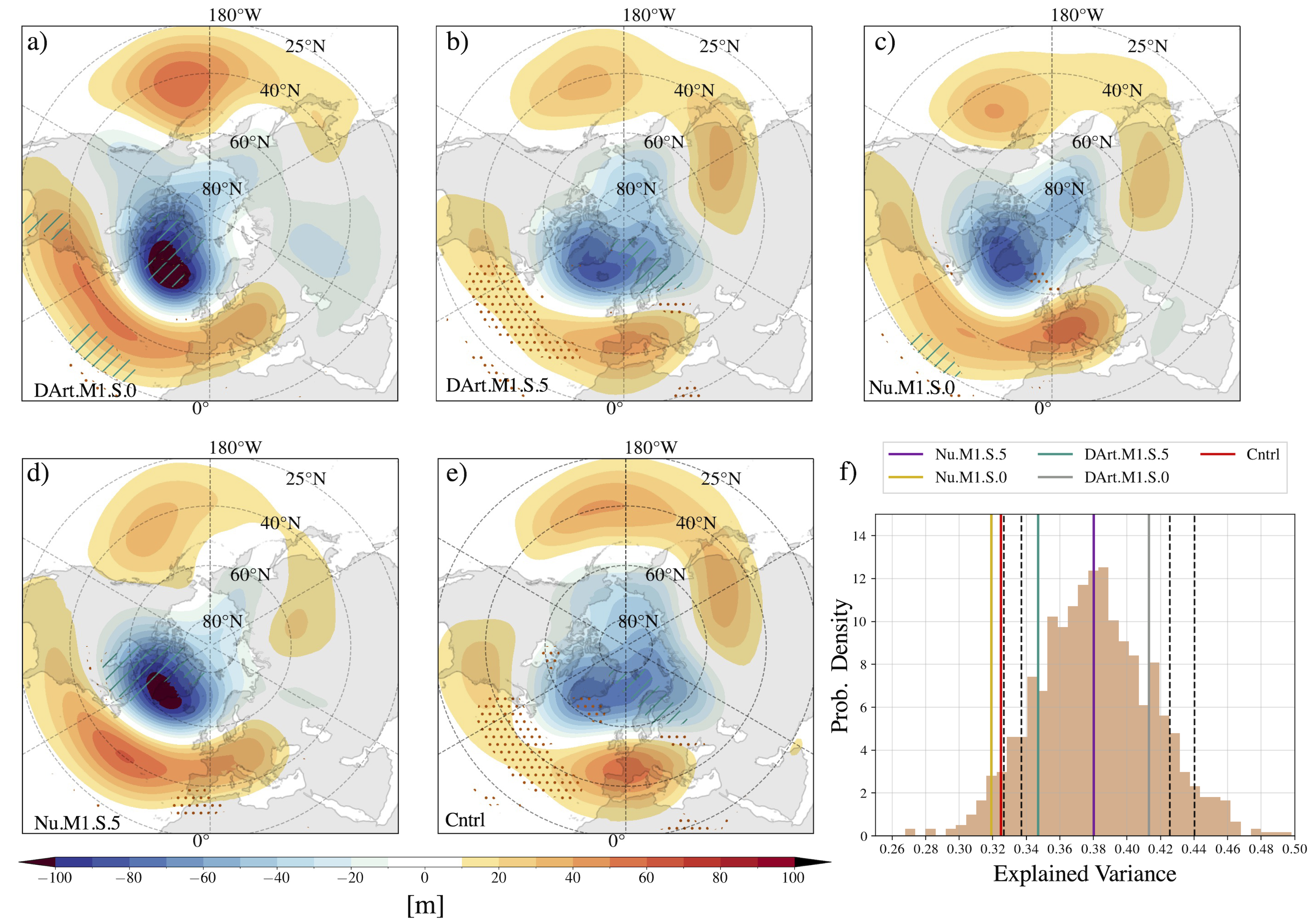}
\caption{As in Fig.9 but for the NAO over the region [20-80°N, 90°W-40°E]
}
\label{fig10}
\end{figure}

Next, we investigate the impacts on the North Atlantic Oscillation, which is the dominant pattern of zonal-mean flow variability across the Atlantic region \citep[][]{DeWeaver2000,Hurrell2003}. The NAO is a North/South dipole characterized by the high latitude, Subpolar Low (SL), and the midlatitude Azores High (Fig. \ref{fig10}). As many studies \citep[][]{Feldstein2003,Franzke2005} show that the NAO is initially driven by upper-level transient eddy vorticity fluxes (which we have shown to improve with MITA/SITA, Fig. 8) we base the NAO pattern on geopotential heights in 500hPa. Focusing first on the Azore's High, we see that the Cntrl simulation underestimates the variance of this pressure pattern across the North Atlantic. This leads to a stunted pattern with two shallowed nodes centered on the eastern United States and over France (as indicated by the stippling). It also has too little variance over the Atlantic sector with an explained variance below the 5th percentile of explained variance in the bootstrapped distribution of observations variances (Fig. \ref{fig10}f). In experiment DArt.M1.S.5 the pattern is unchanged compared to the Cntrl, but the explained variance is significantly increased. Every other simulation results in a significant improvement to the variability over the Atlantic domain. Nu.M1.S.5, DArt.M1S0, DArt.M1.S.5 all have  explained variances that are not significantly different from ERAi (Fig. \ref{fig10}f), while NU.M1.S.0 does not increase the total explained variance over the Cntrl simulation. Additionally, experiments DArt.M1.S0 and Nu.M1.S.5 tend to insert bias over the high latitude SL portion of the NAO. 

In summary, our results indicate that by improving the background state and high-frequency transients, the MITA/SITA increments the NAO variability is improved over the Cntrl experiment. 

\subsection{Streamfunction tendency}

Next, we turn to the evaluation of the intraseasonal growth and decay of the NAO and PNA patterns to assess their realistic representation within the models. First, we define the time-series of each pattern as the monthly (DJF) 500 hPa streamfunction anomaly EOF, projected onto the daily averaged cosine-latitude weighted streamfunction, in the respective NAO and PNA region. The projected time-series is then normalized by its standard deviation. Days with an amplitude of more than 1.5 are defined as PNA/NAO events. As in previous studies, the peak event day, referred to as day 0, is defined as the day on which the magnitude of the index is larger than both the preceding and following days \citep[][]{Franzke2011,Shen2023,Xu2020}. For our definition, the peak day must fall into DJF, though the lagged days prior and after the event can occur in the neighboring months.

\begin{figure}[bt]
\centering
\includegraphics[width=14cm]{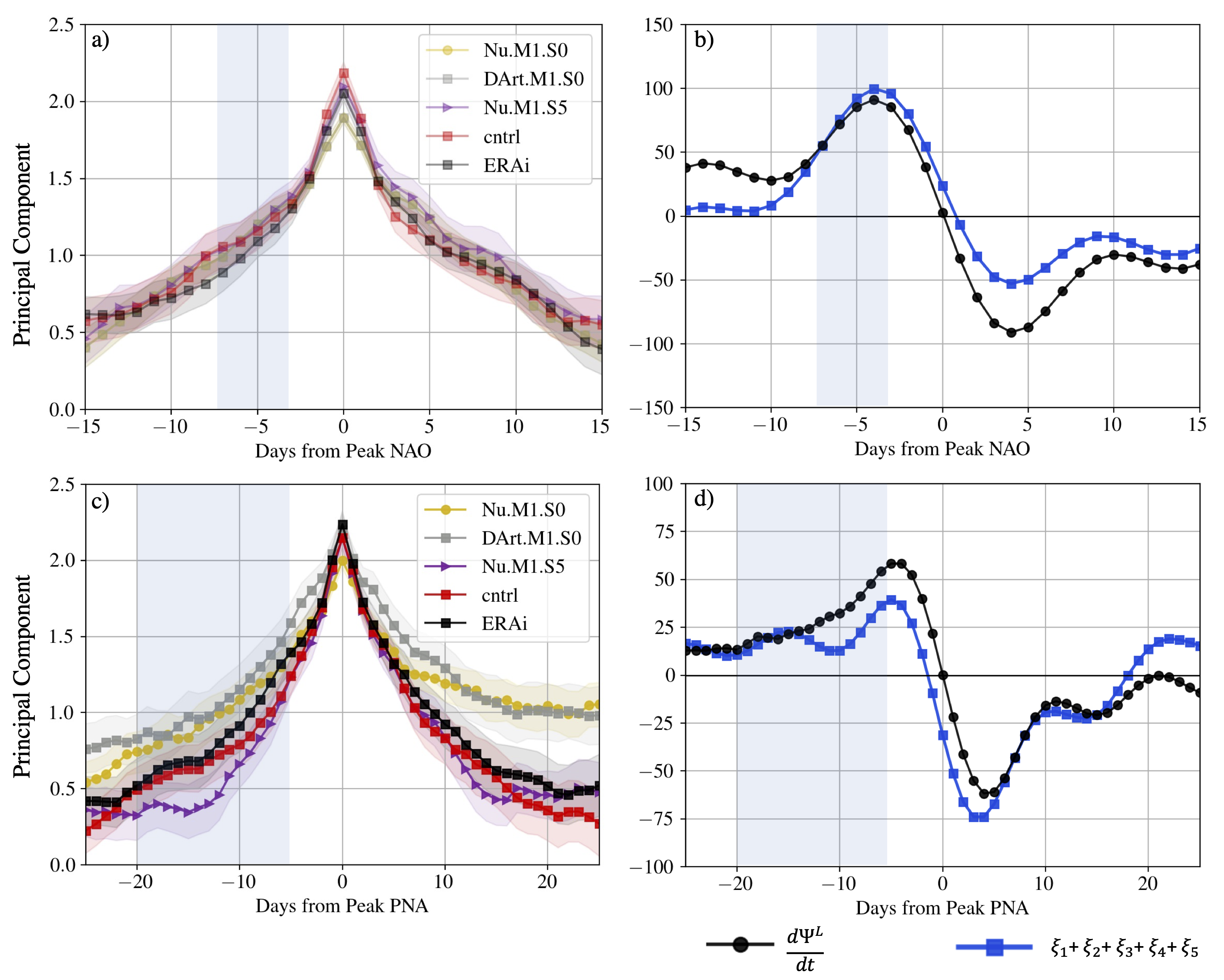}
\captionsetup{}
\caption{(a,c) time evolution of the composited normalized PC time series for the positive NAO (a) and PNA (c) event, the shading indicates the synthetic 80\textsuperscript{th} and 20\textsuperscript{th} percentile of uncertainty via bootstrapping with resampling. Projections for the positive NAO (b) and PNA (d) pattern on the rhs of eq. 3. onto the composited 500-hPa streamfunction anomalies at day M = 0, for \(\sum_{i=1}^{5}\xi_{i}\) (blue line) and \(\frac{\partial\psi^{L}}{\partial t}\) (black line). The ordinate is non-dimensional and has been multiplied by 10\textsuperscript{7}.
}
\label{fig11}
\end{figure}

The lagged composite of the principal component for the positive phase of the NAO and PNA events is shown in Fig. \ref{fig11}. The full cycle of the of the NAO and PNA takes about 3-4 weeks to complete with an e-folding time of $\sim$5-9 days which is consistent with previous studies \citep[e.g.,][]{Cash2001,Dole1990,Franzke2011,Johnson2010}. The NAO growth and decay (Fig \ref{fig11}a) of all experiments is within the uncertainty of ERAi.

This, however, is not the case for the PNA. The addition of the MITA tendency adjustments (DArt.M1.S0 and Nu.M1.S0) alone acts to increase the growth and decay persistence with significant separation from ERAi, Cntrl, and SITA experiments, which form a separate group. This indicates that the bias observed in Fig \ref{fig9} arises from not only the magnitude of the pattern but also the pattern persistence. The addition of the stochastic increments (Nu.M1.S.5) acts to dampen the growth phase somewhat. While it is not significant at the 80th percentile, the mean shows separation in this early growth phase (lag day -20 to -10). The decay phase of Nu.M1.S.5 is similar to those of Cntrl and ERAi.

By comparing the projection of the R.H.S of (3) onto \(\psi_{M}\left(\lambda,\theta\right)\) (blue) to the projections of \(\frac{\partial\psi^{L}}{\partial t}\) (black) onto \(\psi_{M}\left(\lambda,\theta\right)\), we can evaluate the size of the residual, or the error, in the streamfunction decomposition (Fig. \ref{fig11}b, \ref{fig11}d). We show this to validate the accuracy of the given streamfunction decomposition given in the following figures. The projections are formed in the region {[}20-85N, 120E-120W{]} and {[}20-80N, 90W-40E{]} for the PNA and NAO, respectively. These errors likely arise from the interpolation from sigma-pressure hybrid coordinates to pressure coordinates and the use of daily data for the time-differencing rather that data at every reanalysis/model time-step. Here, we selected growth stages with minimal error, namely, the PNA/NAO anomaly growth for the lags denoted by the blue shaded regions in Fig.\ref{fig11}b, d.

\subsection{NAO Anomaly Growth}

\begin{figure}[H]
\centering
\includegraphics[width=14cm]{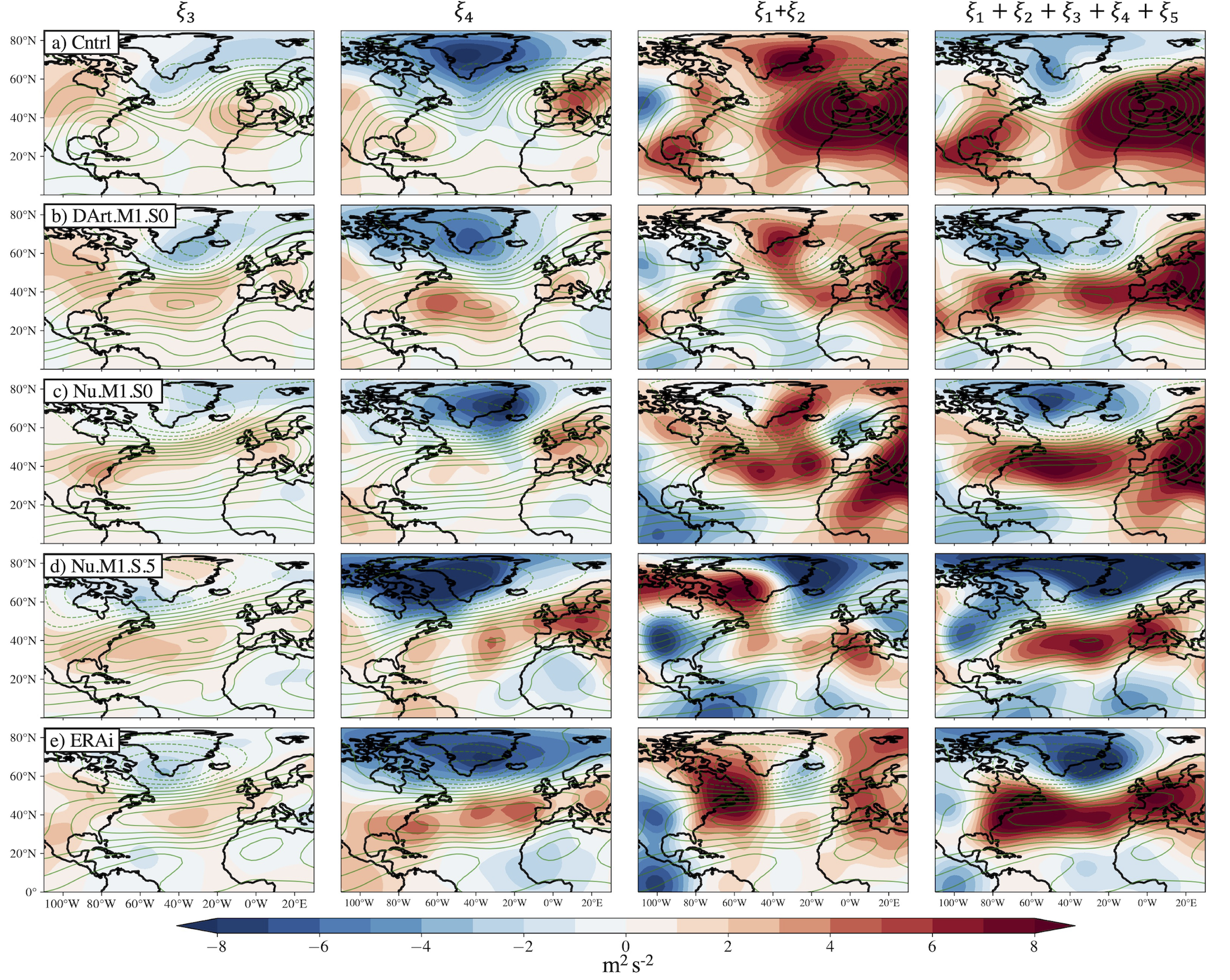}
\captionsetup{}
\caption{Composite lagged projections of the forcing term against the NAO [day M=0] at lag days -7 to -4 for each experiment (a-e, shown in label), column I: \(\xi_{3}\) (high-frequency transients), column II: \(\xi_{4}\) (low-frequency transients), column III: \(\xi_{1}\)+\(\xi_{2}\) (linear terms) and column IV: \(\xi_{1}\)+\(\xi_{2}\)+\(\xi_{3}\)+\(\xi_{4}\)+\(\xi_{5}\) (total estimated streamfunction tendency). The contour (green) shows the -7 to -4 day composite of the low-frequency streamfunction for each experiment.
}
\label{fig12}
\end{figure}

Fig. \ref{fig12} shows the composite streamfunction tendency terms at lag -7 to -4 for the NAO low-frequency tendencies for terms high-frequency (\(\xi_{3}\)), low-frequency transients (\(\xi_{4}\)), the linear terms (\(\xi_{1}\)+\(\xi_{2}\)), and the total streamfunction estimation (\(\xi_{1}\)+\(\xi_{2}\)+\(\xi_{3}\)+\(\xi_{4}\)+\(\xi_{5}\)) in the four examined experiments (Fig. \ref{fig12}a-\ref{fig12}d) and ERAi (Fig. \ref{fig12}e). The composite low-frequency streamfunction pattern for the same period is shown in green contour. Note that we exclude DArt.M1.S.5 as the streamfunction tendency is not significantly different from the Cntrl. For ERAi, the anomalies that comprise the NAO pattern arise more than 7 days before the event peak (Fig. \ref{fig11}b) and the high- and low-frequency transient tendency project most strongly onto the NAO pattern and are dominantly responsible for its growth (Fig.\ref{fig12}). These findings agree with many previous studies \citep[e.g.,][]{Feldstein2003,Franzke2005}. The low-frequency transients (\(\xi_{4}\)) are almost entirely responsible for the growth of the SL (Fig. \ref{fig12}e). Whereas it is the combined effect of the low- and high-frequency transient terms that build the Azores high.

In the Cntrl simulation, it is clear, the linear component (\(\xi_{1}\)+\(\xi_{2}\)) is dominantly responsible for the growth of the two nodal points of the NAO, additionally the linear terms act to dampen the Greenland node of the NAO pattern, which is shown to be biased (Fig. \ref{fig10}e). The linear components of the Cntrl simulation are in stark contrast with the ERAi which does not project strongly onto the NAO, especially over Europe. In summary the lack of variance observed in the Cntrl Azore's High (Fig. \ref{fig9}e), is a result of the lack of high- and low-frequency transient forcing over the Atlantic.

The MITA/SITA simulations show a more physically realistic NAO development, likely meaning that adjusting the background state enables a more realistic propagation of the transients which help to develop the growth of the NAO pattern. We note that every experiment shows an improvement to the upper-level meridional transient transport (Fig. \ref{fig8}), and climatological zonal wind bias (Fig. \ref{fig4}), which lends credence to the improved dynamics based on the growth mechanism of the NAO. The subpolar low is biased in the nudging simulations (Fig. \ref{fig10}), and the tendency decomposition shows that this is manifested through an increase to the low-frequency transient forcing in the model. Our results show that the addition of the MITA/SITA increments can improve not only the background bias but also helps improve the physical representation of the low-frequency NAO streamfunction, and dynamically this occurs through a minimization of bias in the transient forcing.

\subsection{PNA Anomaly Growth}

For studying the growth and maintenance of the PNA the high-frequency (\(\xi_{3}\)) and low-frequency transients (\(\xi_{4}\)) are of less importance. Previous work \citep[][]{Franzke2011,feldstein2002fundamental,Franzke2005} has established that the dominant growth is associated with the stationary eddy advection (eqn. 4), which can be derived from a further decomposition of the linear stream function tendency terms (\(\xi_{1}\)+\(\xi_{2}\)) in equation (3) , (i.e., \(\xi_{1}\)+\(\xi_{2}=\ \xi_{6}+T,\ \) where $T$ represents the remaining forcing terms in the linear portion of the streamfunction decomposition).

The functional form of the stationary eddy advection \(\xi_{6}\) is:

\begin{equation} \label{eq.4}
\xi_{6} = \nabla^{-2}(-{\overline{\mathbf{V}^*}} \cdot \nabla\zeta^{L} - \mathbf{V}^{L} \cdot \nabla[{\overline{\zeta^*}}])
\end{equation}

Where the asterisk denotes deviations from the zonal averages (removing the zonal mean from the climatology). This term represents the induced tendency due to the interaction of the low-frequency transients and the zonally asymmetric time-mean flow. This interaction with the asymmetric background flow has long been studied extensively and shown to be the primary driver of PNA anomalies, anchoring the Aleutian low to the jet exit region.

\begin{figure}[H]
\centering
\includegraphics[width=8cm]{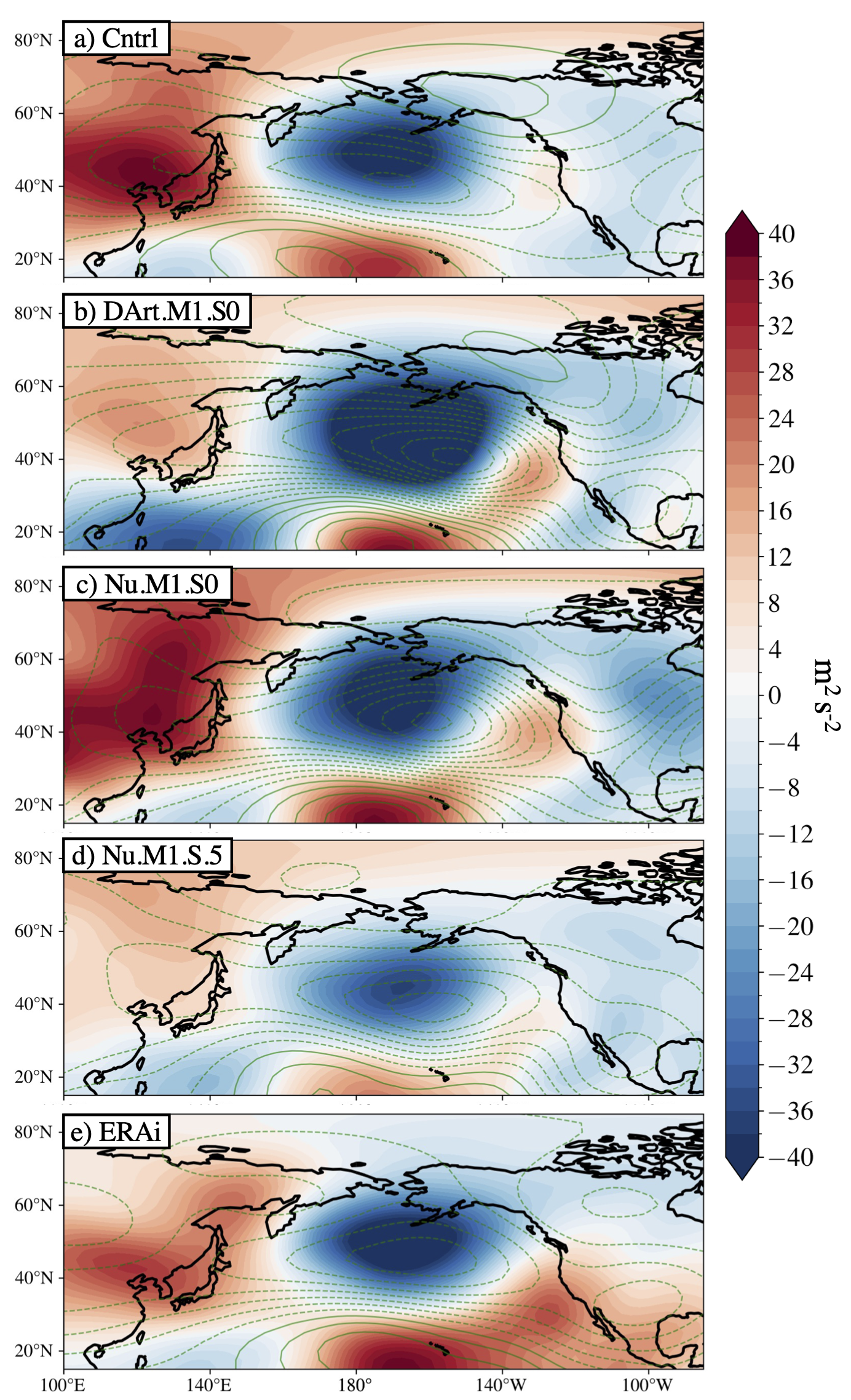}
\captionsetup{}
\caption{Composite lagged projections of the interaction of the transients with the asymmetrical climatological flow streamfunction tendency forcing (\(\xi_{7}\)) against the PNA {[}day M=0{]} at lag days -20 to -5 for each experiment (a-e, shown in label).
}
\label{fig13}
\end{figure}

From Fig. \ref{fig13} we see that Cntrl captures this term quite well when compared to the ERAi, while the MITA experiments (DArt.M1.S0 and Nu.M1.S0) over-energize the stationary eddy advection leading to a deepening of the Aleutian low. By comparing Figs. \ref{fig9} and \ref{fig13} we see that this term indeed controls the bias shown in the Aleutian low region. The addition of the stochastic tendencies (Nu.M1.S.5) reduces the stationary eddy advection, aiding decreased bias in the growth of this pattern.

A further decomposition of \(\xi_{6}\) indicates that it is a change to the low-frequency transients and not an adjustment of the climatology that most dominantly affects the tendency growth (not shown). Previous work \citep[][]{Franzke2011} also showed the importance of the high-frequency transients (\(\xi_{3}\)) in growing and maintaining low-frequency PNA patterns. We find some effects of \(\xi_{3}\) during PNA growth phases (not shown); however, it is much less in magnitude than the tendencies shown in \(\xi_{6}\).

\subsection{Blocking}

\begin{figure}[bt]
\centering
\includegraphics[width=14cm]{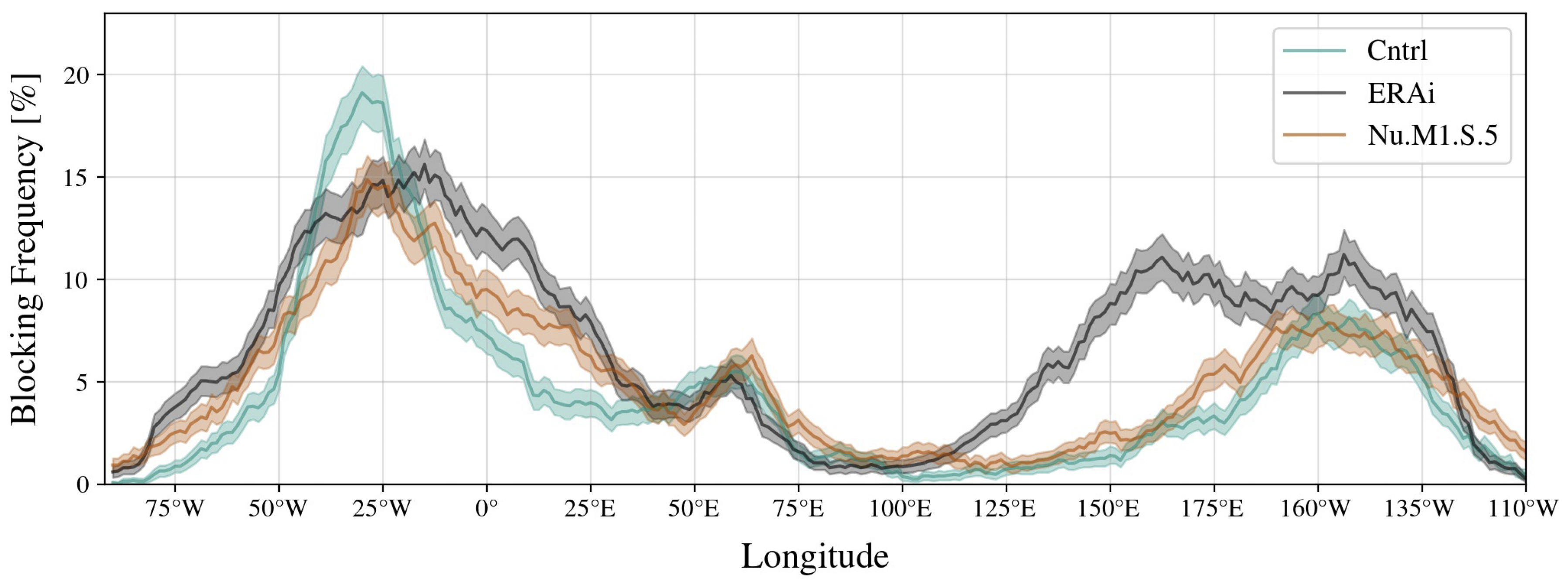}
\captionsetup{font={stretch=1.0}}
\caption{Frequency of DJF Northern Hemisphere blocking events for the ERA-interim reanalysis product (grey), and mode experiments Cntrl (teal) and Nu.M1.S.5 (orange) for the period 1982-2010. The shading indicates the 95\% confidence intervals of the blocking index from each run.
}
\label{fig14}
\end{figure}

Considering the substantial enhancements in addressing climatological biases and capturing low-frequency atmospheric variability during DJF, we evaluate the model adjustment concerning blocking frequency by comparing it to the Cntrl simulation. We focus on a single model experiment, Nu.M1.S.5, as it represents the simulation with the most significant improvements to the NAO. \citet{Kleiner2021} showed that improving a models’ climatological biases can improve the representation of atmospheric blocking events, and it is well documented that many climate models underestimate the observed frequency of blocking events \citep[e.g.,][]{Deandrea1998}. Fig. \ref{fig14} shows the percent of time in DJF where a Z500 hPa block is present at any latitude between 35°N and 65°N in the Cntrl, Nu.M1.S.5, and ERAi from 1982 to 2010. The shaded region shows the 5th and 95th percentile of the bootstrap-estimated uncertainty distribution. Both the Cntrl and Nu.M1.S.5 capture the bimodal nature of the blocking index; however, both models drastically underestimate the frequency of blocks across the western Pacific ocean (120°E-180°), which is a frequent blocking region over the Bering strait \citep[see,][]{Kleiner2021}. The Cntrl experiments shows a high blocking bias in the Atlantic sector that is centered on 30°W, which is immediately upstream of the large node in the Azores High in this model’s representation of the NAO over Europe. Additionally, the Cntrl underestimates blocking occurrence over the western portion of the Atlantic Ocean, which is where the model underestimates NAO variability (Fig. \ref{fig10}e). Adding a stochastic parameterization improves the representation of Northern Hemispheric blocking over the Pacific which has been shown in previous studies \citep[e.g.,][]{Berner2012,Jung2005}. Future analysis will be focus on the connection between NAO and blocking, as well as the physical mechanisms by which the Pacific blocking is improved.

\section{Discussion: Nudging vs. Data Assimilation}
In this study, we conducted a comparison of an online-bias correction via mean and stochastic increment tendency adjustments (MITA/SITA, respectively) to the zonal and meridional wind. The increments were derived from the ensemble adjusted Kalman filter ensemble system DART \citep[][]{Raeder2021}, and a linear relaxation (nudging) to ERA-interim observations. Implicitly, nudging also relies on a DA product, namely the reanalysis product ERA-interim (Fig. \ref{fig2}; green box), the creation of which incurs substantial computational resources (Fig. \ref{fig2}; red box). However, a difference is that a reanalysis data-set only needs to be created once and can then be used for various nudging experiments (Fig. \ref{fig2}; yellow box). The computational ease of this method enables agile rerunning, adjusting, and evaluating of parameter choices for nudging experiments, thereby facilitating the optimization of the bias correction algorithm.

The EAKF is a powerful and elegant algorithm, but it comes with great computational cost, which is dominated by two components: the production of an ensemble of model integrations (for a full-dimensional non-linear estimate of the background error), and computation of the filter products. Integrating the ensemble multiplies the cost of the single-model integration used in some simple data assimilation schemes by a factor of N ensemble members (where, N=80 here) \citep[][]{Anderson2001,Raeder2021}, and involves the systematic start and restart of model runs. It's not feasible to directly compare and quantify the disparity in computational costs between generating the EAKF reanalysis product (DART increments) and the nudging increments. This is because the increments were generated on separate machines. Nevertheless, it's worth noting that the nudging runs utilized approximately 2,500 core-hours per simulation year. It's reasonable to conclude that the EAKF process is significantly more computationally intensive, with its model runtime alone being at least O(100) times higher than that of the nudging approach. We stress that these numbers are for CAM-DART reanalysis and are caused in-part by the specifics of CAM, which is developed as a climate model and thus does not optimize initialization and restart capabilities. While the computation cost of DA to nudging runs will differ from model to model and center to center, the fact that nudging runs are cheaper is generalizable.

Here, we treat the ERAi reanalysis product as ground truth, when in fact, there are significant biases within this product. When verifying against other reanalysis products, our results did not change (not shown), suggesting that the differences between reanalysis products play a small role in our context. A limitation of the nudging approach is that the model improvements will be limited by the quality of the reanalysis it is nudged towards. A good example of this is the discrepancy in the model bias of precipitation in the South Pacific convergence zone (SPCZ) in DJF when ERA-interim precipitation or NOAA GPCP precipitation is used as the ground truth. Precipitation observations over the ocean are notoriously difficult to constrain, and we show that by comparing to the ERAi product, the precipitation is vastly improved in the SPCZ. However, using the GPCP, the model appears to overestimate precipitation in the SPCZ. Dynamically, the lower-level winds, and the vertical velocity are almost without bias in this region after the MITA/SITA adjustments (Fig. \ref{fig6}), and this is a primary driver of tropical precipitation. This adjustment is due to the increments encouraging increased convergence across this region. Until there is agreement on the observations, and the link between the dynamics and parameterized precipitation scheme is solidified, precipitation correction and uncertainty will remain tenuous in any nudging scheme.

\section{Summary and Conclusion}

This study was designed to address three primary questions. Now, we will summarize and conclude our findings by directly addressing those questions. 

1. To which degree do model-error estimates from nudging tendencies agree with those from EAKF analysis increments? 

To our knowledge, there is only one study to compare nudging and DA increments\citep[][]{Jung2011}, and this was done in the context of weather forecasting, and simply for error diagnosis. We conducted a comprehensive comparison between nudging and Ensemble Adjustment Kalman Filter (EAKF) increments, ensuring consistency in model tags, boundary conditions, and forcing, while using the same observational period. We find that overall, nudging tendencies and DA increments pick up the same general features of systematic model bias, particularly at lower model levels (Figure \ref{fig1} and Figure S1). A more detailed analysis reveals differences in the variances of the nudging and the DART tendencies (Figures S2a and S2b). In particular, the observational network is imprinted in the DART tendencies, exhibiting larger variance centered around the locations of the high-density observations' networks, such as radiosonde launching sites and flight paths. We hypothesize that the inhomogeneity in the variance associated with the observations has a negative effect when in the stochastic experiments the random component of the tendencies is reinserted (explainin the results of experiment DArt.M1.S.5, Fig. \ref{fig5}).

2. To what extent does re-inserting DA increments and nudging increments during model runtime reduce climatological model bias of the free-running model?

Overall, we find a positive impact of an online model-error representation based on re-inserting DA increments and nudging on the climatological bias (Figs. \ref{fig5}-\ref{fig8}). Our analysis revealed significant improvements in the background climatology for key surface variables, upper- and lower-level prognostic variables, and transient variables when the model-error scheme was either based on DART or nudging increments. Notably, a 20\% enhancement was observed when comparing the DJF GPCP precipitation against the adjusted zonal and meridional winds. Furthermore, the interaction of non-dynamic variables with the corrected wind field led to additional improvements through dynamic and often non-linear processes. Regional divisions and further variables are presented in the supplemental material, reaffirming the overall finding that this online-bias correction method significantly enhances the background climatology.

3. Will representing subgrid-scale uncertainty in online increment corrections via stochasticity help to improve low-frequency modes of variability without degrading climatological bias?

This study places particular emphasis on improving the representation of not only mean climate but also climate variability, which is essential for the practical applicability of any climate model. Upon examination of the low-frequency variability, it is clear that a mean-state adjustment to the tendencies alone (experiments Nu.S1.M0 and DArt.S1.M0) lead to significant biases in variance across the North Pacific in the PNA pattern. The addition of a stochastic tendency (experiments Nu.S1.M.5 and DArt.S1.M.5) corrected this bias and created an accurate representation of the Aleutian low when compared to observations (Fig. \ref{fig9}). We hypothesize that the stochastic perturbations break up a deepened Aleutian low pattern and correct the mean state climatology while maintaining accurate variability. We explain this result via the stochastics influencing the interaction of the transients with the zonally asymmetric climatological background flow in the North Pacific (Fig. \ref{fig13}). The correction necessitates a proper balance between the adjustment of the background flow and the sub-grid variability (provided by the stochastics), if only one is adjusted, we introduce a model bias. The Cntrl run had a relatively good Aleutian low representation, however, we hypothesize that this is yet another manifestation of compensating biases – by neglecting the subgrid-scale variability the model mean state was tuned toward a shallower Aleutian low than consistent with the mean dynamics. If the mean flow is adjusted, the bias becomes evident, but can be remedied by adding a stochastic component. 

Via a streamfunction tendency decomposition, it is shown that correcting the background state of the model can dramatically improve the representation of high- and low-frequency eddies. This forcing leads to a more accurate representation of the North Atlantic Oscillation, as this pattern is dominantly grown and maintained through the feedback of transient eddy fluxes \citep[e.g.,][]{Feldstein2003,Franzke2005,Lorenz2001}. Furthermore, we show the improvement to the high- and low-frequency eddies intermittently breaks up the zonality of the Northern Hemispheric circulation and leads to a better representation of blocking across the North Atlantic (Fig. \ref{fig14}).

A limitation of this study is that while we accounted for the daily and seasonal cycle, we did not account for state-dependence, as done in the context of machine learning by \citet{Chen2022,WattMeyer2021} and \citet{bonavita2020machine}. This will be the focus of future work together with an assessment of online bias corrections in coupled climate simulations with an evolving ocean. An encouraging result is the improvement of the zonal wind stress, since it has been linked to biases in ocean sea-surface temperatures \citep[][]{Neelin1994}. 

In conclusion, we find that the nudging increment adjustment outperforms the correction provided by the DART increments. A disadvantage of the DA increments is that they depend on observations which are spatially inhomogeneous and can be sparse. Especially in data-limited regions, the analysis increment will unlikely represent model-error. On the other hand, nudging increments will benefit from the balance and conservation properties inherent in reanalysis as well as the spatial homogeneity of a gridded product. A noted disadvantage of the nudging tendencies is that the model will adopt the same biases present in the reanalysis (see our discussion of the SPCZ). While the exact computational cost is system-specific, all data assimilation systems are resource-intensive. Indeed, they are often not readily available, which makes an online-bias correction based on nudging tendencies a viable and flexible approach to study the impact of model bias in climate models.

\section*{Acknowledgements}
This research was made possible by Schmidt Futures, a philanthropic initiative founded by Eric and Wendy Schmidt, as part of its Virtual Earth System Research Institute (VESRI). The CESM project is supported primarily by the National Science Foundation (NSF). This work is supported by the National Center for Atmospheric Research, which is a major facility sponsored by the National Science Foundation under Cooperative Agreement (1852977). We thank Dr. Peiqiang Xu for his useful discussions regarding the streamfunction tendency budget. We thank Dr. Kevin Raeder, Dr. Jefferey Anderson, Dr. Julio Bacmeister, and Dr. Patrick Callaghan for useful discussions regarding data assimilation and nudging.

\section*{Data Availability Statement}
The datasets have been archived in long-term campaign storage at the National Center for Atmospheric Research (NCAR) at data path (\url{/glade/campaign/cisl/aiml/wchapman/CAM_runs/}). It is freely available, but an NCAR registration is required. All code to produce the figures and tables, from the archived record can be found at the authors’ public GitHub repository (\url{https://github.com/WillyChap/MITA_SITA_CAM6}). Additionally, source modifications, model build scripts, and namelists to run the described CAM6 versions can be found in the same repository.

\section*{Supplemental}

Supplemental material is hosted at the authors' public \href{https://github.com/WillyChap/MITA_SITA_CAM6}{GitHub repository}.

\section{Appendix}

\subsection{EOFs}

To extract the leading patterns of variability we perform empirical orthogonal function (EOF) decomposition on the monthly anomaly fields. The climatology is defined as the monthly mean from the full 29-year run or reanalysis product. All EOF patterns are area weighted by the square root of the cosine (latitude) prior to decomposition. We express the orthogonal spatial field as the pointwise regression of each time series with a one-standard deviation change of the temporal principal component. The Pacific - North American Pattern (PNA) and North Atlantic Oscillation (NAO) were examined in detail. These patterns are defined as in Phillips et al. (2014) (NCAR’s Climate Variability Diagnostic Package) as the leading mode of atmospheric variability in the region [20-85°N, 120°E-120°W] and [20-80°N, 90°W-40°E], respectively. 

\subsection{Blocking}

The Scherrer et al., (2006) index, a 2-dimensional extension of the Tibaldi \& Molteni, (1990) index, was employed to identify atmospheric blocking events across a range of central latitudes. The index employs the 500-hPa geopotential height field to detect blocks within the latitudinal band from 35°N to 65°N. The index algorithm applies two criteria to each grid point, namely, (i) a southward displacement of the 500-hPa geopotential height field (Z500) by at least 15° of latitude, indicative of a jet stream reversal, and (ii) a northward excursion of the 500-hPa geopotential height field by at least 150 m at a latitude 15° north of the block, indicative of a meandering jet stream around the blocked area. The criteria must be satisfied for a minimum of 5 consecutive days for an event to qualify as a blocking event.

\subsection{Global Bias Calculation}

As in the NCAR Atmospheric Modeling Working Group Diagnostic Package \citep[][]{AMWG2022}, bias is calculated as the sum of the cosine latitude weighted, root-mean-squared error (RMSE) of the spatial field after a seasonal, monthly, or daily mean has been computed. RMSE was used so that opposite signed local biases do not lead to canceling and erroneously inflated skill.

\end{spacing}
\bibliography{sample}
\selectlanguage{english}
\FloatBarrier
\end{document}